\documentclass[twocolumn,apj]{aastex631}
\usepackage{amsthm,amsmath,amssymb}
\usepackage{url}
\usepackage{mathrsfs}
\usepackage{hyperref}
\usepackage{mathtools}
\usepackage[figuresleft]{rotating}
\usepackage{float}
\usepackage{multirow}
\usepackage{enumitem}

\shorttitle{Inclination Angles of AGN Accretion Disks}
\shortauthors{Du et al.}
\graphicspath{{./}{figures/}}

\begin{document}
\correspondingauthor{Rong Du}
\email{durong@stu.pku.edu.cn}

\title{The Reliability of Accretion Disk Inclination Derived from X-ray Spectroscopy of Active Galaxies}

\author[0009-0006-6543-6333]{Rong Du}
\affiliation{School of Physics, Peking University, Beijing 100871, China}
\affiliation{Department of Astronomy, School of Physics, Peking University, Beijing 100871, China}

\author[0000-0002-5770-2666]{Yuanze Ding}
\affiliation{Division of Physics, Mathematics and Astronomy, California Institute of Technology, Pasadena, CA 91125, USA}
	
\author[0000-0001-6947-5846]{Luis C. Ho}
\affiliation{Department of Astronomy, School of Physics, Peking University, Beijing 100871, China}
\affiliation{Kavli Institute for Astronomy and Astrophysics, Peking University, Beijing 100871, China}

\author[0000-0001-8496-4162]{Ruancun Li}
\affiliation{Department of Astronomy, School of Physics, Peking University, Beijing 100871, China}
\affiliation{Kavli Institute for Astronomy and Astrophysics, Peking University, Beijing 100871, China}

\begin{abstract}
The inclination angle of substructures in active galaxies gives insights into physical components from scales of the vicinity of the central black hole to the entire host galaxy. We use the self-consistent reflection spectral model \textsc{RELXILL} to measure the inclination of the inner region of accretion disks with broadband ($0.3-78\,\rm keV$) X-ray observations, systematically studying the reliability of this methodology. To test the capability of the model to return statistically consistent results, we analyze multi-epoch, joint XMM-Newton and NuSTAR data of the narrow-line Seyfert~1 galaxy I\,Zwicky\,1 and the broad-line radio galaxy 3C\,382, which exhibit different degrees of spectral complexity and reflection features. As expected, we find that adding more data for analysis narrows the confidence interval and that multi-epoch, joint observations return optimal measurements; however, even single-epoch data can be well-fitted if the reflection component is sufficiently dominant. Mock spectra are used to test the capability of \textsc{RELXILL} to recover input parameters from typical single-epoch, joint observations. We find that inclination is well-recovered at 90\% confidence, with improved constraints at higher reflection fraction and higher inclination. Higher iron abundance and corona temperature tighten the constraints as well, but the effect is not as significant as a higher reflection fraction. The spin, however, have little effect in reflection-based inclination measurements. We conclude that broadband reflection spectroscopy can reliably measure inner accretion disk inclination. 
\end{abstract}

\keywords{Accretion (14), Active galaxies (17), X-ray active galactic nuclei (2035), Supermassive black holes (1663)}

\section{Introduction}
\label{sec:Intro}
An active galaxy comprises axisymmetric structures on many scales. On the largest dimensions, the principal axes of the host galaxy of the active galactic nucleus (AGN) are anchored by the global angular momentum distribution of the stars, which in general also defines the orientation of the large-scale interstellar medium. At the opposite extreme, the accretion disk around the central black hole (BH) establishes its orientation based on the angular momentum of its latest fueling episode \citep[e.g.,][]{scheuer1996,king2005,li2015}. Launched by AGNs, the relativistic radio jet that extends from the vicinity of the central BH to possibly beyond the size of the host galaxy is believed to align with the spin axis of the BH (\citealp{blandford1977}; however, see \citealp{natarajan1998}). In the unified model of AGNs, an axisymmetric dusty torus along the equatorial plane of the nucleus obscures the optical or ultraviolet emission from the inner regions \citep{antonucci1993}. The structure and inclination of the torus can be constrained from detailed studies of the infrared spectral energy distribution \citep[e.g.,][]{zhuang2018}.

The fuelling process of AGNs is expected to affect the alignment of the inner accretion disk with respect to the stellar disk of the host galaxy \citep[e.g.,][]{hopkins2012}. However, in practice, the orientation on the smallest scales is observed to be poorly aligned with the host galaxy on larger scales (e.g., \citealp{kinney2000}; see review in Section~3.3.3 of \citealp{kormendy2013}). No preferential alignment exists between AGN jets and the minor axis of elliptical \citep{birkinshaw1985} or disk \citep{wu2022} galaxies. The circumnuclear disks traced by water megamasers also appear misaligned with respect to the host galaxy on global scales \citep[e.g.,][]{greenhill2009,pjanka2017}. Nonetheless, after discarding a minority of highly misaligned sources, \cite{middleton2016} report a loose one-to-one correlation, significant at the $3 \sigma$ level, between the inclination of the galactic stellar disk and the inclination of the AGN accretion disk derived from reflection spectroscopy, hinting on the possibility that AGNs are fed by different mechanisms. 

Radiation generated from the accretion disk photoionizes the broad-line region \citep[e.g.,][]{krolik1998}, whose size and structure can be probed by reverberation mapping experiments \citep{blandford1982}. Dynamical modeling of the broad-line region reveals that its structure is usually axisymmetric (disk-like; e.g., \citealp{pancoast2014,li2018,williams2018}), although its formation mechanism is still not firmly established. Hypotheses linking accretion disk winds to the origin of the broad-line region have been discussed for years \citep[e.g.,][]{emmering1992,murray1995,czerny2017}, with recent attention focusing on the scenario of a failed dusty outflow \citep[e.g.,][]{czerny2011blr,czerny2016,baskin2018}. Cross-matching the inclination of the accretion disk and the inclination of the broad-line region inferred from dynamical modeling can offer valuable insights into the physical relationship between these two fundamental components of the AGN central engine.

Measuring the inner accretion disk inclination itself plays a crucial role in understanding the environment around supermassive BHs and the processes of accretion and outflow. Theoretically, if we accept general relativity \citep{einstein1916}, the no-hair conjecture \citep[e.g.,][pp. 875--877]{misner1973} limits parameters describing the configuration of every Kerr-Newman BH \citep{kerr1963,newman1965} to its mass $M_\mathrm{BH}$, charge $Q$\footnote{Charged BHs are usually not considered in an astrophysical context \citep[however, see e.g.,][]{komissarov2022}.}, and angular momentum $\mathbf{J}$, which can be decomposed further to a dimensionless spin parameter $a_\ast \coloneqq |\mathbf{J}|c/G M_\mathrm{BH}^2$ and two angular components, including the orientation of the BH. The inclination $\theta_\mathrm{disk}$ of the inner accretion disk in AGNs, defined as the viewing angle of the system with respect to its normal, is believed to align with the orientation of the central BH because of Lense-Thirring precession \citep[the Bardeen-Petterson effect:][]{bardeen1975,pringle1992,papaloizou1995}. On a more pragmatic level, knowing the inner disk inclination helps to improve estimates of the intrinsic bolometric luminosity of the AGN, assuming that the inner disk aligns with the outer disk emitting in other wavelengths\footnote{This is not always true \citep[e.g., see simulations of][]{liska2021}.}. Disk inclination measurements also help determine the intrinsic values of projected quantities, such as radial velocities. Precise measurements of inclinations would open up the possibility of employing inclination-dependent theoretical models in data analysis. As an instance in which X-ray spectroscopy utilizes the forward-folding technique to minimize the difference between the observed spectra and an instrument-convolved model, fitting the thermal blackbody continuum of a relativistic, optically thick, geometrically thin accretion disk (\citealp{shakura1973}; extended in the relativistic regime by \citealp{novikov1973}) with a weak X-ray corona gives loose constraints on BH spin, whose accuracy would be improved with knowledge of the inclination \citep[e.g.,][]{czerny2011,done2013,reynolds2021}. 

Various efforts have been made to measure the inclination angle of AGN accretion disks. For sources that have a jet, an approximate orientation of the beam, which is equal to the inner disk inclination, can be readily estimated from the radio core dominance parameter \citep[e.g.,][]{ghisellini1993,wills1995} via very-long-baseline interferometry measurements of the fastest part of the jet. Other possible approaches include fitting the profile of double-peaked broad H$\alpha$ emission lines with a relativistic disk model (\citealp{eracleous1994,storchibergmann1995}; e.g., see \citealp{storchibergmann1997} for NGC\,1097 and \citealp{ho2000} for NGC\,4450), studying $\mathrm{H_{2}O}$ megamaser disks \citep[e.g., in NGC\,4258; ][]{herrnstein1999}, reproducing near-infrared and mid-infrared photometric and interferometric observations \citep[e.g., for NGC\,1068;][]{honig2007}, and measuring the configuration of circumnuclear gas with high-spatial resolution, near-infrared spectroscopy \citep[e.g., for NGC\,3227, NGC\,4151, and NGC\,7469;][]{hicks2008}. In works that assume the unified model of AGNs, the inner disk inclination is deemed to roughly equal the inclination of other substructures, for example, the narrow-line region \citep{fischer2013,fischer2014} and broad-line region \citep{wu2001,zhang2002} clouds. 

Another approach to derive key disk properties of AGNs, including inclination, is by fitting the reflection components of their broadband X-ray spectrum \citep[e.g.,][]{fabian1989}. Empirically, the X-ray emission in AGNs shows similar properties, leading to the conventional view that electrons from a hot corona in the vicinity of the BH inverse \cite{compton1923} scatter thermal optical/ultraviolet photons from the accretion disk into the X-rays, creating a power-law continuum with an exponential cutoff \citep{thorne1975,haardt1991,haardt1993,dove1997,belmont2008}. Apart from this power-law continuum, observations of AGNs also identify two distinct features commonly depicted as the consequence of coronal X-rays reflected by a standard thin disk \citep{novikov1973,shakura1973}: one is the Fe~K$\alpha$ line at $6-7\,\rm keV$ \citep{fabian1989,tanaka1995}, the other the Compton hump peaking at $20-40\,\rm keV$ \citep{guilbert1988,lightman1988,matt1991}. The former originates from X-ray fluorescence, while the latter traces continuum emission Compton-scattered into the shape of a hump, with the range determined by photoelectric absorption on the red edge and Compton recoil, \cite{klein1929} cross-section, and cutoff temperature on the blue side. As the irradiation process mainly occurs in the inner disk region, gravitational redshift, relativistic beaming, and the Doppler effect modify the emission lines, which stretch the low-energy wing and create a sharper blueshifted peak. In addition, the relativistically smeared reflection fluorescent lines in the soft band may be a possible origin of the soft excess \citep[e.g.,][]{fabian2002}. On measuring the inclination, the key physical process is the broadening of the Fe~K$\alpha$ profile, which offers information on the velocity of the emitting fluid element as a function of disk inclination and the radius of the emitters. The red wing of the Fe line, used to measure the BH spin \citep[e.g.,][]{reynolds2014,reynolds2021}, is sensitive to the inner radius of the emitters, whereas the blue edge is sensitive to the inclination owing to Doppler boosting \citep{rees1966}. Consequently, the inclination in disk reflection models is degenerate with the spin and the emissivity profile. The iron abundance of the disk, to the extent that it modifies the strength of the Fe line, also comes into play. Moreover, according to \citet{fabian2015}, broadband X-ray spectroscopy can place stringent constraints on the geometry and radiative mechanisms of the corona. Thus, the measurement of inclination can be influenced by the coronal temperature and the overall proportion of the reflection component (reflection fraction, $R_f$), which affect the spectrum's flux, the shape of the Compton hump, and the relative strength of the Fe line. 

Numerical models have been developed for calculating reflection spectra. \textsc{PEXRAV} \citep{magdziarz1995}, \textsc{REFLIONX} \citep{ross2005}, and \textsc{XILLVER} \citep{garcia2010,garcia2011,garcia2013} specialize in calculating the intrinsic spectrum. For relativistic smearing, kernels like \textsc{DISKLINE} \citep{fabian1989}, \textsc{LAOR} \citep{laor1991}, \textsc{KYRLINE} \citep{dovciak2004}, \textsc{KERRDISK} \citep{brenneman2006}, and \textsc{RELLINE} \citep{dauser2010,dauser2013} have been readily applied. Advanced reflection models that calculate broadband relativistic reflection continuum include the \textsc{KY} package \citep{dovciak2004,dovciak2022}, \textsc{REFLKERR} \citep{niedzwiecki2019}, \textsc{RELTRANS} \citep{ingram2019,mastroserio2021}, and \textsc{RELXILL} \citep{dauser2014,garcia2014}. As a self-consistent, angle-dependent scheme, \textsc{RELXILL} is developed by point-wise convolving \textsc{XILLVER} with \textsc{RELLINE}. The former model derives the reflected radiation field from a thin slab using atomic calculations from \textsc{XSTAR} \citep{bautista2001,kallman2001}, and the latter relativistically blurs the output of \textsc{XILLVER} using a general relativistic ray-tracing technique.

Our larger aim, the subject of forthcoming works in this series, is to explore the connection between the inner accretion disk inclination and other facets of the AGN and its host galaxy. As an initial step, this paper focuses on ascertaining the degree to which the \textsc{RELXILL} reflection model can deliver reliable measurements of the inner disk inclination under conditions that mimic observations of interest. Systematic uncertainties need to be understood given the complexity of the model and the degeneracy of model parameters. Our method is explained in Section~\ref{sec:method}. Section~\ref{sec:obs_fitting} describes the tests and results on observed spectra, and Section~\ref{sec:simulate} presents the analysis of mock spectra. Section~\ref{sec:discussion} discusses the sensitivity of the inclination measurements to the reflection fraction, iron abundance, coronal temperature, model selection, and some technical concerns about binning. Conclusions appear in Section~\ref{sec:conclusion}. We adopt a cosmology with $H_0 = 70\, \rm km\,s^{-1}\,Mpc^{-1}$, $\Omega_\Lambda = 0.73$ and $\Omega_m = 0.27$.

\section{Methodology}
\label{sec:method}

\subsection{Systematic Inclination Measurement}
\label{sec:systematical_method}
We systematically measure inner accretion disk inclination by analyzing X-ray spectra with \textsc{RELXILL} as a baseline model. Instead of rigidly fitting the spectra by brute-force computation, we exploit the flexibility of \textsc{RELXILL} to evaluate judiciously and systematically models with an increasing number of emission or absorption components (e.g., soft excess and ionized absorbers). The fit jointly solves for $\theta_\mathrm{disk}$ with other parameters, including the spin $a_\ast$, the iron abundance $A_\mathrm{Fe}$, the power-law photon index $\Gamma$, the ionization state of the inner disk $\xi$, the electron temperature of the corona $kT_{e}$, the reflection fraction $R_f$, and the radial emissivity profile of the primary source (i.e., the corona). We implement \textsc{RelxillCp} in the \textsc{RELXILL} family to account simultaneously for part of the soft excess, the broad Fe line, and the Compton hump. This choice is motivated by the thermal Comptonization model \citep[\textsc{NTHCOMP};][]{zdziarski1996,zycki1999} of the corona in \textsc{RelxillCp}, which is more physical than the cutoff power-law model employed in, for example, \textsc{RELXILL} and \textsc{RelxillD}. 

To achieve a wide energy range from 0.3 to 78\,keV, we combine the high-quality spectra from the pn charge-coupled device \citep[CCD;][]{struder2001} and the Metal Oxide Semiconductor \citep[MOS;][]{turner2001} of the European Photon Imaging Camera (EPIC) detectors onboard the X-ray Multi-Mirror Mission \cite[XMM-Newton;][]{jansen2001} to capture complex emission and obscuration features in the softer bands, together with spectra from the Nuclear Spectroscopy Telescope Array \cite[NuSTAR;][]{harrison2013} mission, which extends the bandpass coverage to include the Compton hump and offers more information on the Fe line. For basic discussions, we always take advantage of XMM-Newton (pn, MOS1, and MOS2) to provide the high-resolution soft X-ray spectrum of the soft excess and complicated absorption components, which are crucial in broadband fitting. 

We fit the spectra with \textsc{RelxillCp} in the \textsc{RELXILL} family (version 2.2), as implemented in \textsc{Xspec} \cite[12.12.1;][]{arnaud1996}. During the fit, we make the underlying assumption that the ``disk-related'' parameters ($\theta_\mathrm{disk}$, $a_\ast$, and $A_\mathrm{Fe}$), except for ionization $\xi$, do not vary on our timescales of interest, but that the ``corona-related'' parameters ($\Gamma$, $kT_{e}$, $R_f$, and the emissivity parameters, including the two indices Index$_\mathrm{1}$ and Index$_\mathrm{2}$, as well as the broken radius $R_\mathrm{br}$), can vary freely. The fixed parameters are tied among the spectra from different epochs; the rest are untied. For simplicity, we lump $\xi$ together with the set of corona-related parameters hereinafter, although it is actually a disk parameter. Our baseline model is a combination of Galactic absorption, a blackbody soft excess, and \textsc{RelxillCp}\footnote{In the terminology of \textsc{Xspec}: \\ \textsc{constant*TBabs*(zBBody+RelxillCp)}}. For different combinations of data, we adopt different technical strategies. 

We divide the data and corresponding strategy into five cases:

\begin{enumerate}
\item Single-epoch XMM-Newton EPIC. The lack of higher energy observation is the main motivation to use this method. All relevant parameters are set free and fit jointly.

\item Single-epoch XMM-Newton observation and a non-simultaneous NuSTAR observation. We fix $\theta_\mathrm{disk}$, $a_\ast$, and $A_\mathrm{Fe}$, and fit the data jointly while setting $\Gamma$, $\xi$, $kT_\mathrm{e}$, $R_f$, and the emissivity parameters (Index$_\mathrm{1}$, Index$_\mathrm{2}$, $R_\mathrm{br}$) free.

\item Single-epoch XMM-Newton observation and a simultaneous NuSTAR observation. In this fairly ideal case, one that is commonly adopted in recent reflection modeling, we tie both the disk-related and corona-related parameters and set free the cross-calibration constant and normalization parameter.

\item Multi-epoch, simultaneous XMM-Newton and NuSTAR observations, together with additional non-simultaneous NuSTAR data. Where there are multiple epochs of simultaneous XMM-Newton and NuSTAR observations, adding additional NuSTAR data, even if not contemporaneous, may be helpful to improve the fit of the Compton hump because the count rate above 20\,keV is relatively low for typical Seyfert~1 galaxies. Similar to previous cases, we tie relevant parameters within each simultaneous epoch and leave free the cross-calibration terms. The additional data are fitted with extra corona-related parameters, except that the disk-related parameters are tied for all spectra.

\item Multi-epoch, simultaneous XMM-Newton and NuSTAR data. We assume that all parameters in the same epoch are the same but may vary between different epochs.
\end{enumerate}

\noindent 
In all five cases, the blackbody temperature $kT$ is a free parameter but assumed constant within all the spectral epochs, while the Galactic absorption column density is fixed to the value retrieved from the Leiden/Argentine/Bonn Galactic H\,{\small I} Survey \citep{kalberla2005}. Note that within each simultaneous epoch, the parameters for the spectra from different detectors are tied. Cross-calibration constants ($\sim 1$) account for the differences between instruments, and normalization parameters are used to compensate for differences in integrated flux and exposure time. Upon achieving the best-fit parameters, we generate a posterior parameter distribution to determine their confidence intervals.

\subsection{Reliability Tests}
\label{sec:reliability_method}

In this section, we address the challenges inherent in the \textsc{RELXILL} methodology and outline our design for reliability tests. Despite its self-consistent integration of physical parameters, the model's high dimensionality and nonlinearity pose challenges, particularly in retrieving global minima during spectral fitting. This complexity introduces intricacies in interpreting results, compounded by the multitude of parameter degeneracies involved.

To establish the reliability of our methodology, we systematically discuss three aspects:
\begin{enumerate}
    \item Statistical consistency: We scrutinize the consistency of model fits across datasets with varying quality and quantity, with a primary focus on statistical errors during observed spectra fitting. By analyzing model performance under diverse data conditions, we identify cases that minimize errors. Insights and technical strategies to enhance inclination measurements are provided based on broadband fits to the observed spectra of I\,Zwicky\,1 and 3C\,382 under various combinations.

    \item Computational robustness: Assessing the robustness and accuracy of parameter retrieval, we acknowledge potential biases introduced by the effectiveness of the model coding and the theoretical explainability of the model. Parameter retrieval is evaluated through tests on mock spectra simulated with known input parameters, offering insights into model performance under idealized conditions. Our emphasis in this work diverges from previous studies adopting similar approaches \citep{bonson2016, choudhury2017, kammoun2018}.

    \item Theoretical generalizability: We explore the model's capacity to explain observed data and underlying physics across different scenarios. Although generalizability primarily relates to the model's development, observers play a crucial role in selecting an appropriate model. We provide a detailed rationale for choosing \textsc{RelxillCp} in Section~\ref{sec:relselect}.
\end{enumerate}

For a complex model such as this, parameter space exploration and error calculation are non-trivial. We employ a Markov chain Monte Carlo (MCMC) algorithm \citep{metropolis1953} to generate the posterior distribution of baseline parameters. We check for chain convergence and use the 90\% confidence interval to document the results of spectral fits, facilitating the assessment of the consistency of $\theta_\mathrm{disk}$ measurements across various circumstances.

\section{Fitting Observed Spectra}
\label{sec:obs_fitting}

\begin{deluxetable}{cccc}[t]
\label{tab:izw_spec_divid}
\tablenum{1}
\setcounter{table}{1}
\caption{Epochs of I\,Zwicky\,1 Observations}
\tablehead{
\colhead{Epoch} &
\colhead{Observatory} &
\colhead{Obs. ID} &
\colhead{Effective Exposure}\\
\ &&&\colhead{(ks)}
}
\startdata
$a$ & NuSTAR & 60501030002 & 53.3 \\ \hline
$b$ & \begin{tabular}[c]{@{}c@{}}XMM-Newton\\ NuSTAR\end{tabular} & \begin{tabular}[c]{@{}c@{}}0851990101\\ 60501030002\end{tabular} & \begin{tabular}[c]{@{}c@{}}50.3 \\ 38.2\end{tabular} \\ \hline
$c$ & NuSTAR & 60501030002 & 40.2\\ \hline
$d$ & \begin{tabular}[c]{@{}c@{}}XMM-Newton\\ NuSTAR\end{tabular} & \begin{tabular}[c]{@{}c@{}}0851990201\\ 60501030002\end{tabular} & \begin{tabular}[c]{@{}c@{}}47.2\\ 35.9 \end{tabular} \\ \hline
$e$ & NuSTAR & 60501030002 & 63.9
\enddata
\end{deluxetable}

\begin{deluxetable}{cccc}[t]
\label{tab:3c_spec_divid}
\tablenum{2}
\setcounter{table}{2}
\caption{Epochs of 3C\,382 Observations}
\tablehead{
\colhead{Epoch} &
\colhead{Observatory} &
\colhead{Obs. ID} &
\colhead{Effective Exposure}\\
\ &&&\colhead{(ks)}
}
\startdata
$a$ & \begin{tabular}[c]{@{}c@{}}XMM-Newton\\ NuSTAR\end{tabular} & \begin{tabular}[c]{@{}c@{}}0790600101\\ 60202015002\end{tabular} & \begin{tabular}[c]{@{}c@{}}18.8\\ 23.0\end{tabular} \\ \hline
$b$ & \begin{tabular}[c]{@{}c@{}}XMM-Newton\\ NuSTAR\end{tabular} & \begin{tabular}[c]{@{}c@{}}0790600201\\ 60202015004\end{tabular} & \begin{tabular}[c]{@{}c@{}}19.0\\ 24.6\end{tabular} \\ \hline
$c$ & \begin{tabular}[c]{@{}c@{}}XMM-Newton\\ NuSTAR\end{tabular} & \begin{tabular}[c]{@{}c@{}}0790600301\\ 60202015006\end{tabular} & \begin{tabular}[c]{@{}c@{}}21.7\\ 20.8\end{tabular} \\ \hline
$d$ & \begin{tabular}[c]{@{}c@{}}XMM-Newton\\ NuSTAR\end{tabular} & \begin{tabular}[c]{@{}c@{}}0790600401\\ 60202015008\end{tabular} & \begin{tabular}[c]{@{}c@{}}19.0\\ 21.7\end{tabular} \\ \hline
$e$ & \begin{tabular}[c]{@{}c@{}}XMM-Newton\\ NuSTAR\end{tabular} & \begin{tabular}[c]{@{}c@{}}0790600501\\ 60202015010\end{tabular} & \begin{tabular}[c]{@{}c@{}}16.4\\ 21.0\end{tabular}
\enddata
\end{deluxetable}

\begin{figure*}[ht]
\centering
\includegraphics[width=0.9\linewidth]{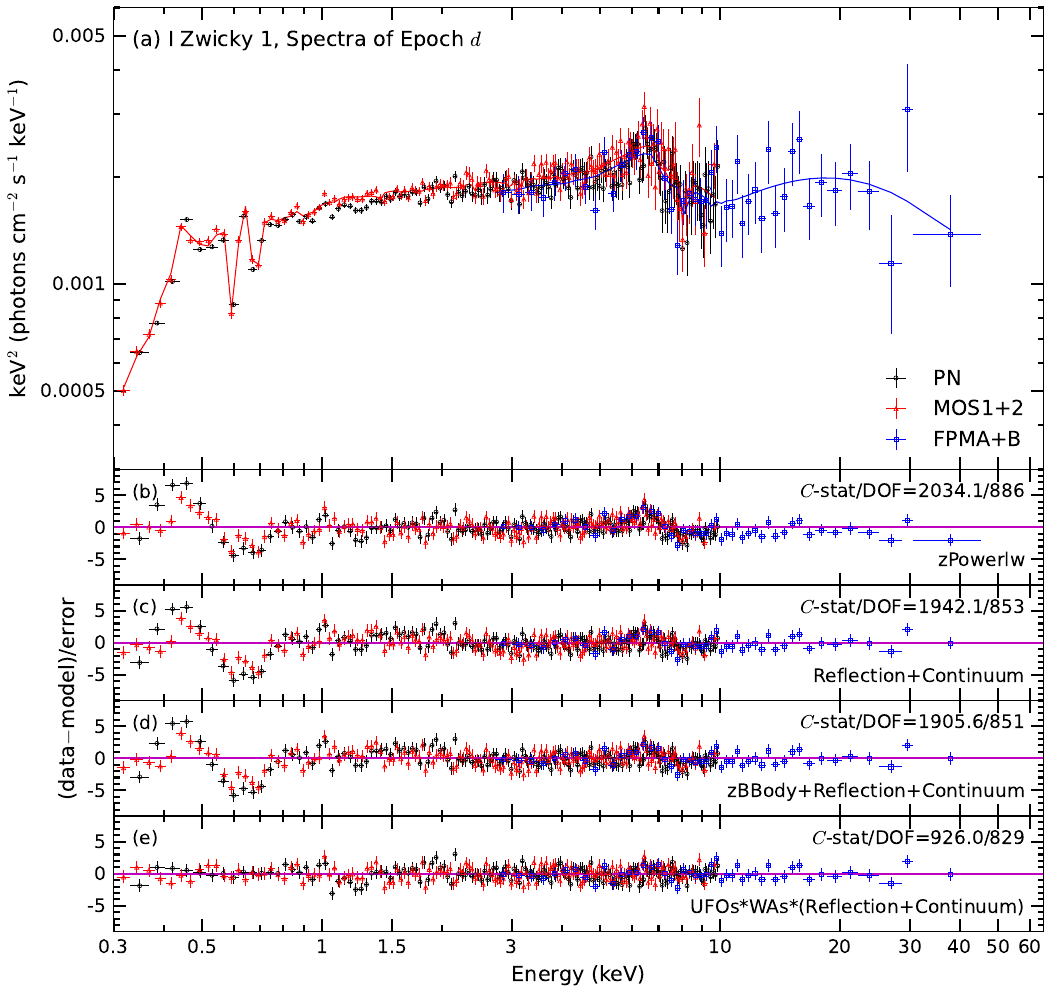}
\caption{(a) The XMM-Newton and NuSTAR broadband spectral fit of I\,Zwicky\,1, showing the best fit that contains the Galactic absorbed reflection model, warm absorbers (WAs), and ultra-fast outflows (UFOs). For clarity, only the spectra of epoch~$d$ are shown. Note that the spectra of all five epochs are fitted simultaneously. The residuals in terms of sigmas with error bars of size one are shown for various models: (b) the phenomenological redshifted power-law model, (c) \textsc{RelxillCp}, (d) \textsc{RelxillCp} plus a redshifted blackbody component to represent a mild soft excess, and (e) the best-fit model, with ionized absorbers added to the model of panel (c).}
\label{fig:izw_spec}
\end{figure*}

\begin{figure*}[ht]
\centering
\includegraphics[width=0.9\linewidth]{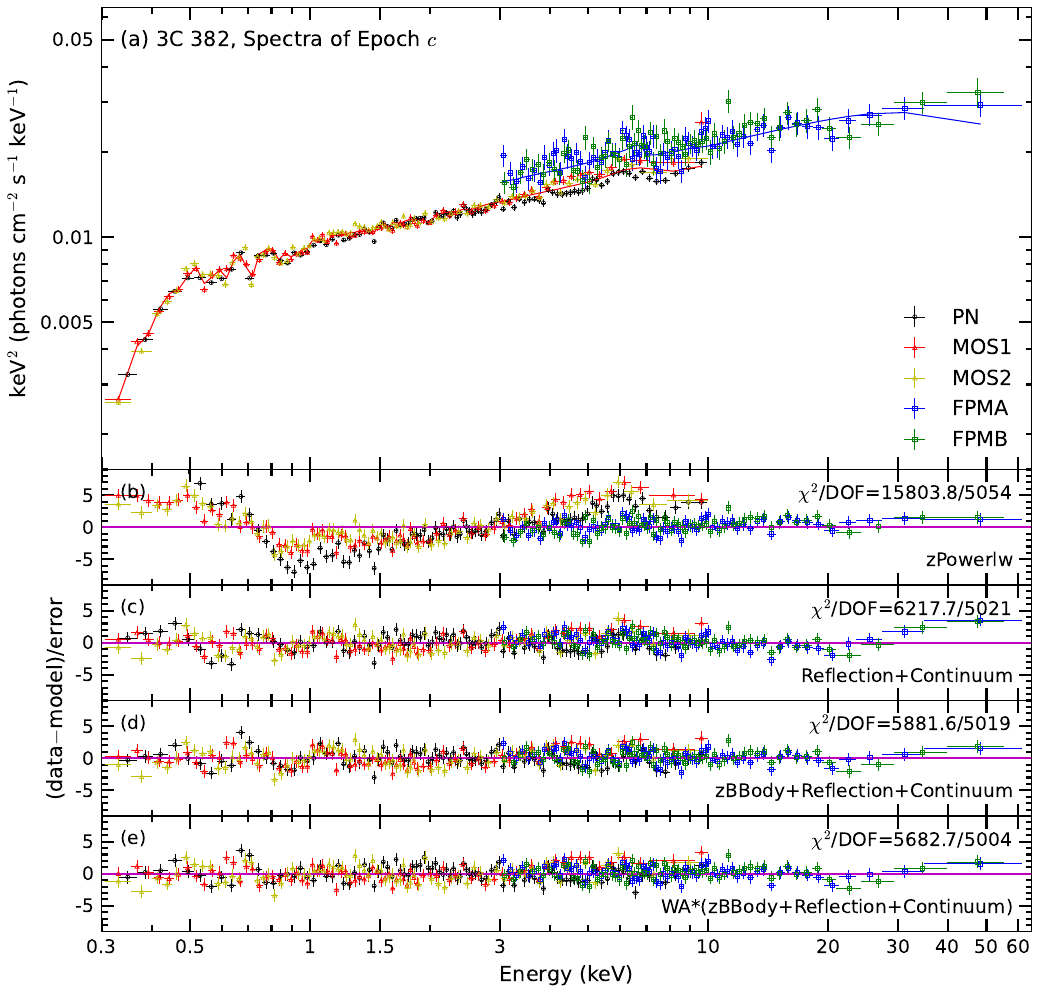}
\caption{(a) The XMM-Newton and NuSTAR broadband spectral fit of 3C\,382, showing the best fit that contains the baseline model and a warm absorber (WA). For clarity, only the spectra of epoch~$c$ are shown. For plotting purposes, the XMM-Newton data have been rebinned to $\rm S/N > 30$, while the NuSTAR data have been rebinned to $\rm S/N > 10$. The spectra of five epochs are fitted simultaneously. The residuals in terms of sigmas with error bars of size one are shown for various models: (b) the phenomenological redshifted power-law model, (c) \textsc{RelxillCp}, (d) \textsc{RelxillCp} plus a redshifted blackbody component to represent a mild soft excess, and (e) the best-fit model, with a WA added to the previous model.}
\label{fig:3c_spec}
\end{figure*}

\begin{deluxetable*}{ccccccccccc}[t]
\tablenum{3}
\caption{Best-fit Parameters for the Continuum of I\,Zwicky\,1}
\setcounter{table}{3}
\tablehead{
\colhead{Epoch} &
\colhead{$\theta_\mathrm{disk}$} &
\colhead{$a_\ast$} &
\colhead{$A_\mathrm{Fe}$} &
\colhead{$R_f$} &
\colhead{$kT_{e}$} &
\colhead{$\Gamma$} &
\colhead{Index$_\mathrm{1}$} &
\colhead{Index$_\mathrm{2}$} & 
\colhead{$R_\mathrm{br}$} &
\colhead{$\log \xi$}\\
\ &\colhead{($^\circ$)}&&&& \colhead{(keV)} &&&&\colhead{($r_{g}$)}&
\colhead{($\mathrm{erg~cm~s^{-1}}$)}\\
\colhead{(1)} &
\colhead{(2)} &
\colhead{(3)} &
\colhead{(4)} &
\colhead{(5)} &
\colhead{(6)} &
\colhead{(7)} &
\colhead{(8)} &
\colhead{(9)} &
\colhead{(10)} &
\colhead{(11)}
}
\startdata
$a$& & & & $>8.95$ & $49.2_{-8.8}^{+9.8}$ & $2.10_{-0.07}^{+0.06}$ & $7.64_{-0.63}^{+0.77}$ & $3.00_{-0.37}^{+0.28}$ & $2.8 \pm 0.4$ & $3.27_{-0.10}^{+0.14}$ \\
$b$& & & & $1.93_{-0.31}^{+0.42}$ & $85.9_{-23.1}^{+22.8}$ & $2.13 \pm 0.01$ & $>8.63$ & $2.96_{-0.48}^{+0.36}$ & $2.4_{-0.2}^{+0.5}$ & $3.47_{-0.11}^{+0.09}$ \\
$c$& $44.3_{-2.4}^{+2.8}$ & $>0.978$ & $3.01 \pm 0.34$ & $3.46 \pm 0.33$ & $293_{-40}^{+65}$ & $2.09_{-0.07}^{+0.06}$ & $8.81_{-0.77}^{+0.67}$ & $3.00_{-0.25}^{+0.31}$ & $2.0_{-0.6}^{+0.3}$ & $3.47_{-0.16}^{+0.21}$ \\
$d$& & & & $3.66_{-0.37}^{+0.69}$ & $17.5_{-1.4}^{+1.2}$ & $1.88 \pm 0.01$ & $>8.83$ & $2.98_{-0.34}^{+0.23}$ & $2.5_{-0.2}^{+0.3}$ & $3.37 \pm 0.08$ \\
$e$& & & & $2.10_{-0.49}^{+0.72}$ & $134_{-10}^{+6}$ & $2.15_{-0.06}^{+0.05}$ & $6.16_{-1.13}^{+1.03}$ & $2.95_{-0.63}^{+0.33}$ & $2.1 \pm 0.3$ & $3.31_{-0.12}^{+0.17}$
\enddata
\tablecomments{
Col. (1): Epoch.
Col. (2): Inclination angle of the inner accretion disk. 
Col. (3): Dimensionless BH spin.
Col. (4): Iron abundance.
Col. (5): Reflection fraction.
Col. (6): Temperature of the corona.
Col. (7): Photon index.
Col. (8): Power-law index 1 for the broken power-law disk emissivity.
Col. (9): Power-law index 2 for the broken power-law disk emissivity.
Col. (10): Break radius for the broken power-law disk emissivity, in units of the gravitational radius, $r_{g}=GM/c^{2}$.
Col. (11): Ionization parameter of the accretion disk. 
The spin, inclination, and iron abundance are fitted simultaneously for the five epochs, instead of only for epoch~$c$.}
\label{tab:izw1_reflection_par}
\end{deluxetable*}

\begin{deluxetable*}{ccccccccccc}[t]
\tablenum{4}
\caption{Best-fit Parameters for the Continuum of 3C\,382}
\setcounter{table}{4}
\tablehead{
\colhead{Epoch} &
\colhead{$\theta_\mathrm{disk}$} &
\colhead{$a_\ast$} &
\colhead{$A_\mathrm{Fe}$} &
\colhead{$R_f$} &
\colhead{$kT_{e}$} &
\colhead{$\Gamma$} &
\colhead{Index$_\mathrm{1}$} &
\colhead{Index$_\mathrm{2}$} & 
\colhead{$R_\mathrm{br}$} &
\colhead{$\log \xi$} \\
\ &\colhead{($^\circ$)}&&&& \colhead{(keV)} &&&&\colhead{($r_{g}$)}&
\colhead{($\mathrm{erg~cm~s^{-1}}$)} \\
\colhead{(1)} &
\colhead{(2)} &
\colhead{(3)} &
\colhead{(4)} &
\colhead{(5)} &
\colhead{(6)} &
\colhead{(7)} &
\colhead{(8)} &
\colhead{(9)} &
\colhead{(10)} &
\colhead{(11)} 
}
\startdata
$a$&&&& $0.24_{-0.06}^{+0.05}$ & $22.1_{-3.2}^{+4.9}$ & $1.75 \pm 0.01$ & $5.15_{-0.53}^{+0.43}$ & $0.11_{-0.02}^{+0.05}$ & $23.7_{-3.9}^{+5.2}$& $3.10 \pm 0.22 $ \\
$b$&&&& $0.34_{-0.07}^{+0.08}$ & $15.6_{-1.5}^{+2.3}$ & $1.72 \pm 0.01$ & $5.65_{-0.42}^{+0.46}$ & $1.32_{-0.28}^{+0.33}$ & $10.4_{-2.2}^{+2.1}$ & $3.44 \pm 0.13$ \\
$c$&$41.8_{-4.1}^{+2.7}$&$> 0.992$ & $>8.19$& $0.49_{-0.13}^{+0.10}$ & $23.0_{-3.0}^{+5.8}$ & $1.79 \pm 0.01$ & $7.79_{-1.65}^{+1.26}$ &$3.00_{-0.29}^{+0.40}$ & $2.7_{-0.5}^{+0.4}$ & $2.90_{-0.10}^{+0.12}$ \\
$d$&&&& $0.50_{-0.12}^{+0.07}$ & $16.8_{-1.8}^{+3.0}$ & $1.78 \pm 0.01$ & $5.86_{-0.41}^{+0.35}$ & $1.06_{-0.14}^{+0.15}$ & $12.0_{-1.8}^{+2.5}$ & $2.79_{-0.10}^{+0.12}$ \\
$e$&&&&$0.55_{-0.10}^{+0.05}$ & $29.7_{-5.2}^{+7.5}$ &  $1.80_{-0.02}^{+0.01}$ & $8.38_{-0.80}^{+0.51}$ & $1.89_{-0.25}^{+0.34}$ & $4.6_{-0.5}^{+0.6}$ & $ 3.02_{-0.16}^{+0.14}$ \\
\enddata
\tablecomments{
Col. (1): Epoch.
Col. (2): Inclination angle of the inner accretion disk. 
Col. (3): Dimensionless BH spin.
Col. (4): Iron abundance.
Col. (5): Reflection fraction.
Col. (6): Temperature of the corona.
Col. (7): Photon index.
Col. (8): Power-law index 1 for the broken power-law disk emissivity.
Col. (9): Power-law index 2 for the broken power-law disk emissivity.
Col. (10): Break radius for the broken power-law disk emissivity, in units of the gravitational radius, $r_{g}=GM/c^{2}$.
Col. (11): Ionization parameter of the accretion disk. 
The spin, inclination, and iron abundance are fitted simultaneously for the five epochs, instead of only for epoch~$c$.}
\label{tab:3c_reflection_par}
\end{deluxetable*}

\subsection{The Sources}
\label{sec:sample}
The prototypical narrow-line Seyfert~1 galaxy I\,Zwicky\,1 has a complicated soft X-ray absorption structure and a strong Fe~K$\alpha$ line \citep{ding2022}, which is thought to be evidence of strong relativistic disk reflection \cite[however, see][]{reeves2019}. The warm absorbers (WAs) in the soft X-rays, a combination of fast outflows and corresponding shocked circumnuclear gas \citep{ding2022}, evolve over long timescales. Changes in ionization may be driven by the different densities of the clumps crossing the observer’s line-of-sight, in which the ``skin'' layer facing the source has a higher ionization state \citep{costantini2007,silva2018}. I\,Zwicky\,1 was observed once in 2020 by NuSTAR in a $\sim5$ day-long exposure and twice by XMM-Newton, each with a $\sim 70\, \rm ks$ exposure (PI: D. R. Wilkins). Because of the strong variability, we divide the NuSTAR data into five epochs according to their overlap with the XMM-Newton observations \citep[as in][see our Table~\ref{tab:izw_spec_divid}]{ding2022}. 

The broad-line radio galaxy 3C\,382 exhibits a WA, a soft excess, and an Fe~K$\alpha$ line, but no Compton hump \citep{torresi2010,ballantyne2014}. According to \citet{ursini2018}, the soft excess comes from a warm Comptonization component with a photon index $\Gamma \approx 2.4-2.5$, a temperature of 0.6\,keV, and a corresponding optical depth of $\sim 20$. These parameters are all in general agreement with those in radio-quiet Seyfert galaxies \citep[e.g.,][]{jin2012,petrucci2018}. The Fe~K$\alpha$ line is broad in two epochs and possibly broad in the other three, which, together with the lack of a Compton hump, seemingly undermines the reflection model previously applied to explain the properties of the source \citep[e.g.,][]{sambruna2011}. However, given that the lower limit of the coronal temperature may be as low as $20-40\,\rm keV$ and the reflection fraction is merely $\sim 0.1$ \citep{ursini2018}, the reflection model is intrinsically poorly constrained, and additional information is needed to test it. For our purposes, the broadening of the iron line is a sufficient indicator of reflection-based inclination measurements. Compared to I\,Zwicky\,1, 3C\,382 has weaker reflection features and less complex outflow structures. It was observed simultaneously five times in 2016 (PI: F. Ursini; Table~\ref{tab:3c_spec_divid}).

\subsection{Data Reduction} 
\label{sec:data_reduc}
For XMM-Newton data, we extract the light curve and spectrum from the pn and MOS of the EPIC detectors with the System Analysis Software (version 20.0.0) and the calibration files released on 31 March 2022. For each bundle of observational data, we use \texttt{epproc} and \texttt{emproc} to retrieve its event lists, \texttt{evselect} to extract the spectrum of the source and background, and \texttt{arfgen} and \texttt{rmfgen} to generate, respectively, the ancillary files and redistribution matrices. Checking for pile-up and background flares reveals that no corrections are needed for either source. The spectra are extracted from a circular region with a 40$''$ diameter around the source. In the case of the pn observations, a source-free polygon region excluding serendipitous sources and chip edges is chosen for background extraction, whereas for MOS observations the background spectrum is extracted from a 300$''$-diameter circular region within the range of the outer CCD. For I\,Zwicky\,1, we combine the spectra from the MOS1 and MOS2 detectors with the task \texttt{epiccombine} to enhance the signal-to-noise ratio (S/N). We rebin the spectra of 3C\,382 with \texttt{specgroup} so as to not oversample the full width at half-maximum resolution by more than a factor of 3, and to ensure that each spectral bin contains at least 25 counts. 

The NuSTAR data are reduced using \textsc{NuSTARDAS} version 2.1.2, with event lists retrieved from the two Focal Plane Module (FPM) detectors processed by the standard \texttt{NUPIPELINE}, which is integrated into the \textsc{HEASoft} bundle (6.30.1). We use calibration files from \textsc{CALDB} version 20220802. A source-free 300$''$-diameter circle near the source region is used to measure the background. For I\,Zwicky\,1, we extract the spectra and light curve with a small circular region of 30$''$ diameter centered on the point source, as recommended for faint sources, and then combine the data from FPMA and FPMB following \citet{wilkins2021}. For 3C\,382, to stay consistent with XMM-Newton procedures, we also use a circular region of 40$''$ diameter centered on the point source to extract the spectrum. The source spectra of 3C\,382 are further rebinned with \texttt{ftgrouppha} in \textsc{HEASoft} such that each spectral bin contains at least 25 counts. 

\begin{deluxetable}{ccccc}[ht]
\tablenum{5}
\caption{Best-fit Parameters for Ionized Absorbers of I\,Zwicky\,1}
\setcounter{table}{5}
\tablehead{
\colhead{Epoch} &
\colhead{IA} &
\colhead{$N_\mathrm{H}$} &
\colhead{$\log \xi$} &
\colhead{$v/c$} \\
& & \colhead{($10^{20}\,\rm cm^{-2}$)} & 
\colhead{($\mathrm{erg~cm~s^{-1}}$)} & \\
\colhead{(1)} &
\colhead{(2)} &
\colhead{(3)} &
\colhead{(4)} &
\colhead{(5)} 
}
\startdata
\multirow{4}{*}{$b$} & UFO1 & $2.09^{+0.25}_{-0.31}$ & $-1.72^{+0.14}_{-0.12}$ & $0.34 \pm 0.01$\\ 
& UFO2 & $7.07^{+0.87}_{-0.82}$ & $0.85^{+0.12}_{-0.08}$ & $0.112^{+0.0008}_{-0.0013}$\\
& WA1 & $2.46 \pm 0.29$ & $1.82^{+0.21}_{-0.27}$ & $0.035 \pm 0.004$ \\
& WA2 & $6.16^{+0.37}_{-0.33}$ & $0.695 \pm 0.063$ & $0.044 \pm 0.001$ \\
\hline
$c$& WA3 & $4790^{+590}_{-410}$ & $3.44^{+0.16}_{-0.36}$ & $0.25^{+0.02}_{-0.03}$ \\ \hline
\multirow{5}{*}{$d$} & UFO1 & $2.45 \pm 0.21$ & $-0.711_{-0.037}^{+0.035}$ & $0.276_{-0.010}^{+0.008}$ \\ 
& UFO2 & $3.44^{+0.32}_{-0.37}$ & $0.283^{+0.026}_{-0.025}$ & $0.124^{+0.009}_{-0.006}$ \\
& WA1 & $1.38^{+0.11}_{-0.07}$ & $1.034^{+0.047}_{-0.051}$ & $0.036^{+0.003}_{-0.002}$ \\
& WA2 & $2.42_{-0.21}^{+0.32}$ & $1.21_{-0.08}^{+0.11}$ & $0.046 \pm 0.001$ \\
& WA3 & $495^{+47}_{-39}$ & $3.64^{+0.14}_{-0.32}$ & $0.22^{+0.02}_{-0.03}$ \\
\hline
$e$& WA3 & $2270 \pm 180$ & $3.94^{+0.36}_{-0.46}$ & $0.27^{+0.02}_{-0.04}$
\enddata
\tablecomments{
Col. (1): Epoch.
Col. (2): Ionized absorber (IA) component, including two ultra-fast outflows (UFO1, UFO2) and three warm absorbers (WA1, WA2, WA3).
Col. (3): Hydrogen column density.
Col. (4): Ionization parameter.
Col. (5): Outflow velocity divided by the speed of light.}
\label{tab:izw1_wa_par}
\end{deluxetable}

\begin{deluxetable}{cccc}[ht]
\tablenum{6}
\caption{Best-fit Parameters for Ionized Absorbers of 3C\,382}
\setcounter{table}{6}
\tablehead{
\colhead{Epoch} &
\colhead{$N_\mathrm{H}$} &
\colhead{$\log \xi$} &
\colhead{$v/c$} \\
\ & \colhead{($10^{20}\,\rm cm^{-2}$)} & 
\colhead{($\mathrm{erg~cm~s^{-1}}$)}
\\
\colhead{(1)} &
\colhead{(2)} &
\colhead{(3)} &
\colhead{(4)} 
}
\startdata
$a$& $16.1_{-3.2}^{+2.8}$ & $1.92 \pm 0.15$ & $-0.132_{-0.019}^{+0.009}$ \\
$b$& $357_{-79}^{+80}$ & $2.88_{-0.04}^{+0.03}$ & $0.007_{-0.006}^{+0.002}$ \\
$c$& $23.5_{-4.9}^{+4.7}$ & $1.68_{-0.08}^{+0.10}$ & $-0.125_{-0.010}^{+0.019}$ \\
$d$& $55.2_{-10.8}^{+25.3}$ & $2.51_{-0.11}^{+0.13}$ & $-0.013_{-0.008}^{+0.010}$ \\
$e$& $25.9_{-4.5}^{+3.2}$ & $2.02_{-0.07}^{+0.0.05}$ & $0.006 \pm 0.008$ \\
\enddata
\tablecomments{
Col. (1): Epoch.
Col. (2): Hydrogen column density.
Col. (3): Ionization parameter.
Col. (4): Outflow velocity divided by the speed of light.}
\label{tab:3c_wa_par}
\end{deluxetable}

\begin{deluxetable*}{cccccc}[ht]
\tablenum{7}
\caption{Luminosity, Flux, and Best-fit Parameters for Cross-calibration Constants of I\,Zwicky\,1}
\setcounter{table}{7}
\tablehead{
\colhead{Epoch} &
\colhead{$L_\mathrm{0.1-200\,keV}$} &
\colhead{$F_\mathrm{0.3-2\,keV}$} &
\colhead{$F_\mathrm{2-10\,keV}$} &
\colhead{$F_\textsc{RELXILL}$} &
\colhead{\textsc{constant}} \\
\ & 
\colhead{($10^{44}\,\rm erg~s^{-1}$)} & 
\colhead{($10^{-11}\,\rm erg~cm^{-2}~s^{-1}$)} &
\colhead{($10^{-11}\,\rm erg~cm^{-2}~s^{-1}$)} &
\colhead{($10^{-11}\,\rm erg~cm^{-2}~s^{-1}$)} & 
\colhead{MOS/PN} \\
\colhead{(1)} &
\colhead{(2)} &
\colhead{(3)} &
\colhead{(4)} &
\colhead{(5)} &
\colhead{(6)}
}
\startdata
$a$& $0.71 \pm 0.02$ & $0.06 \pm 0.01$ & $0.39 \pm 0.01$ & $0.78 \pm 0.02$ \\
$b$& $1.01 \pm 0.01$ & $0.53 \pm 0.01$ & $0.48 \pm 0.01$ & $1.03 \pm 0.01$ & $1.02 \pm 0.01$\\
$c$& $1.19 \pm 0.03$ & $0.09 \pm 0.01$ & $0.62 \pm 0.01$ & $1.26 \pm 0.03$ \\
$d$& $0.94 \pm 0.01$ & $0.40 \pm 0.01$ & $0.51 \pm 0.01$ & $0.96 \pm 0.01$ & $1.03 \pm 0.01$\\
$e$& $1.06 \pm 0.02$ & $0.09 \pm 0.01$ & $0.53 \pm 0.01$ & $1.12 \pm 0.02$ \\
\enddata
\tablecomments{
Col. (1): Epoch.
Col. (2): $0.1-200\, \rm keV$ luminosity of the model.
Col. (3): $0.3-2\, \rm keV$ flux of the data.
Col. (4): $2-10\, \rm keV$ flux of the data.
Col. (5): $0.3-78\, \rm keV$ flux of the \textsc{RelxillCp} component.
Col. (6): Cross-calibration constants of epochs~$b$ and $d$; we set the \textsc{constant} of PN and FPM to 1 and show the results of the MOS spectra.
}
\label{tab:izw1_flux_lum}
\end{deluxetable*}

\begin{deluxetable*}{cccccc}[ht]
\tablenum{8}
\caption{Luminosity, Flux, and Best-fit Parameters for Cross-calibration Constants of 3C\,382}
\setcounter{table}{8}
\tablehead{
\colhead{Epoch} &
\colhead{$L_\mathrm{0.1-200\,keV}$} &
\colhead{$F_\mathrm{0.3-2\,keV}$} &
\colhead{$F_\mathrm{2-10\,keV}$} &
\colhead{$F_\textsc{RELXILL}$} &
\colhead{\textsc{constant}} \\
\ & 
\colhead{($10^{44}\,\rm erg~s^{-1}$)} & 
\colhead{($10^{-11}\,\rm erg~cm^{-2}~s^{-1}$)} &
\colhead{($10^{-11}\,\rm erg~cm^{-2}~s^{-1}$)} &
\colhead{($10^{-11}\,\rm erg~cm^{-2}~s^{-1}$)}  & 
\colhead{MOS1,2/PN; FPMB/A} \\
\colhead{(1)} &
\colhead{(2)} &
\colhead{(3)} &
\colhead{(4)} &
\colhead{(5)} &
\colhead{(6)}
}
\startdata
$a$& $9.35 \pm 0.12$ & $0.41 \pm 0.01$ & $3.96 \pm 0.05$ & $11.69 \pm 0.15$ & $1.00 \pm 0.01$, $1.02 \pm 0.01$; $1.00 \pm 0.03$ \\
$b$& $9.73 \pm 0.11$ & $0.47 \pm 0.01$ & $4.54 \pm 0.05$ & $12.60 \pm 0.15$ & $1.03 \pm 0.01$, $0.95 \pm 0.01$; $1.01 \pm 0.03$ \\
$c$& $10.35 \pm 0.13$ & $0.49 \pm 0.01$ & $4.61 \pm 0.06$ & $13.03 \pm 0.16$ & $1.02 \pm 0.01$, $1.03 \pm 0.01$; $1.01 \pm 0.03$ \\
$d$& $9.11 \pm 0.12$ & $0.45 \pm 0.01$ & $4.24 \pm 0.05$ & $11.80 \pm 0.15$ & $1.03 \pm 0.01$, $1.02 \pm 0.01$; $1.01 \pm 0.02$ \\
$e$& $10.98 \pm 0.13$ & $0.52 \pm 0.01$ & $4.82 \pm 0.06$ & $13.52 \pm 0.16$ & $1.02 \pm 0.01$, $1.02 \pm 0.01$; $1.04 \pm 0.03$ \\
\enddata
\tablecomments{
Col. (1): Epoch.
Col. (2): $0.1-200\, \rm keV$ luminosity of the model.
Col. (3): $0.3-2\, \rm keV$ flux of the data.
Col. (4): $2-10\, \rm keV$ flux of the data.
Col. (5): $0.3-78\, \rm keV$ flux of the \textsc{RelxillCp} component.
Col. (6): Cross-calibration constants; we set the \textsc{constant} of PN and FPMA to 1, and arrange the results in the sequence of MOS1, MOS2, and FPMB.
}
\label{tab:3c_flux_lum}
\end{deluxetable*}

\subsection{Broadband Model Fitting}
\label{sec:modelfit}
Based on the data rebinning scheme, we apply the \cite{cash1979} statistics ($C$-stat) for I\,Zwicky\,1 during statistics minimization, and the $\chi^2$ statistics for 3C\,382. I\,Zwicky\,1 falls in case~4 and 3C\,382 in case~5 discussed in Section~\ref{sec:systematical_method}. All the epochs of 3C\,382 and epochs~$b$ and $d$ of I\,Zwicky\,1 belong to case~3. We produce a case~4 fit for I\,Zwicky\,1 and a case~5 fit for 3C\,382. We adopt the conventional scheme by starting from a phenomenological power-law model and gradually modify the model components. Because X-ray spectra are folded, fitting with a simple model first allows us to get an initial overall sketch of the spectra. 

We start by fitting the data with a redshifted power-law model (\textsc{zPowerlw}) modified by Galactic absorption \citep[\textsc{TBabs}; see][]{wilms2000}. The broadband spectra of both sources show deviations from the phenomenological model, which require the addition of ionized absorbers at $\sim$0.5\,keV, a broad Fe~K$\alpha$ line, and a high-energy Compton hump, motivating us to use \textsc{RelxillCp}. A mild excess in the residuals of the soft band prompts us to add a redshifted blackbody component (\textsc{zBBody}), but we do not consider intrinsic absorbers as in previous works \citep{ursini2018,ding2022}. As discussed for cases~4 and 5, corona-related parameters are untied for non-simultaneous observations, while disk-related parameters are tied. Furthermore, we fix parameters of the accretion disk that cannot be derived directly from our analysis, including its inner radius ($R_\mathrm{in}$, radius of the innermost stable circular orbit), outer radius ($R_\mathrm{out} = 400\,r_{g}$, with $r_{g} \coloneqq GM/c^2$), and density $n = 10^{15}\,\mathrm{cm^{-3}}$. The broadband model fitting procedure is illustrated in Figure~\ref{fig:izw_spec} for I\,Zwicky\,1 and in Figure~\ref{fig:3c_spec} for 3C\,382\footnote{Figure~\ref{fig:3c_spec}b shows a discrepancy between XMM–Newton and NuSTAR spectra ($\Delta \Gamma \sim 0.1$), as also reported in \citet{ursini2018}. According to \citet{furst2016} and \citet{ponti2018}, NuSTAR sometimes measures a higher flux than XMM-Newton, especially between 3 and 5\,keV, which may lead to steeper FPM spectra. In our case the difference in fluxes is present in the entire overlapping range of 3--10\, keV.}. 

We remark on a technical issue related to the blackbody component in best-fit models. Figure~\ref{fig:izw_spec}e presents the best-fit model for I\,Zwicky\,1 as \textsc{constant*TBabs*XSTAR*RelxillCp}. In contrast to a direct combination of \textsc{XSTAR} tables and the baseline model (with \textsc{zBBody}), we exclude the blackbody component because the successful modeling of complex emission and absorption features in the soft band with \textsc{XSTAR} tables renders the addition of a \textsc{zBBody} inconsequential. In this case, if a blackbody is added to the best-fit model, the reduced $C$-stat changes negligibly ($\Delta C\text{-stat}/\Delta \mathrm{DOF} = -0.1/-2$). Our best-fit model for I\,Zwick\,1 aligns with that of \citet{ding2022}. Yet for 3C\,382 the blackbody component is crucial. As a test, removing the epoch-independent redshifted blackbody component leads to a significant reduction in fit statistics ($\Delta \chi^2/\Delta\mathrm{DOF} = -186.8/-2$). \citet{ursini2018} also emphasized the inadequacy of pure reflection models in capturing the soft excess. However, this epoch-independent treatment might include biases, which will be addressed in Section~\ref{sec:soft_excess}.

Ionized absorbers exist in both sources. Since we focus on interpreting the reflection model, the absorption components are fitted following \citet{ding2022} for I\,Zwicky\,1 and \citet{torresi2010} for 3C\,382. Specifically, we fit the ultra-fast outflows and WAs in I\,Zwicky\,1 with five \textsc{XSTAR} absorbers and five \textsc{XSTAR} additive tables, adopting the parameters of \citet{ding2022}. We fit the WA in 3C\,382 with multiplicative \textsc{XSTAR} tables. The reason for not adopting the two-state model in \citet{ursini2018} is that our analysis does not include Reflection Grating Spectrometer \citep[RGS;][]{den-herder2001} data and may not be able to resolve the two states. In terms of fit statistics over degrees of freedom (DOF), the single-state absorber model returns $\chi^2/\mathrm{DOF} = 5682.6/5004 = 1.136$, whereas the double-state model yields $\chi^2/\mathrm{DOF} = 5641.4/4989 = 1.131 \: (\Delta \chi^2/\Delta\mathrm{DOF} = -41.2/-15)$. The improvement is not significant. 

We generate the posterior probability distribution for each parameter using MCMC implemented in \textsc{XSPEC}. Adopting the \cite{goodman2010} algorithm, we use 200 walkers with a chain of $6\times10^6$ elements ($3\times10^4$ iterations). The proposal distribution is generated from a Gaussian distribution around the best-fit values. The chain convergence is tested using graphs to monitor the evolution of parameters and $\chi^{2}$ with regard to MCMC steps, and with the integrated autocorrelation time $\tau_{f}$ \citep{goodman2010} estimated by the function in \textsc{EMCEE} \citep{foremanmackey2013}. We ensure that the MCMC chains are longer than $\sim 1000 \, \tau_{f}$, as suggested by \citet{sokal1996}. The first $6\times10^5$ elements are rejected as the burn-in phase prior to storing the chain, in order that the chain forgets the initialization point and samples in the equilibrium state. The results of reflection continuum fitting of I\,Zwicky\,1 and 3C\,382 are given in Tables~\ref{tab:izw1_reflection_par} and \ref{tab:3c_reflection_par}, respectively. The best-fit parameters of the ionized absorbers are shown in Table~\ref{tab:izw1_wa_par} for I\,Zwicky\,1 \citep[also see][their Table~2]{ding2022}, and in Table~\ref{tab:3c_wa_par} for 3C\,382. 

Last but not least, Tables~\ref{tab:izw1_flux_lum} and \ref{tab:3c_flux_lum} provide the calculated luminosity and fluxes. Cross-calibration constants $\sim 1$ yield good agreement between the MOS detectors and PN on XMM-Newton, as well as between the FPMA and FPMB on NuSTAR.

\begin{figure*}
\centering
\includegraphics{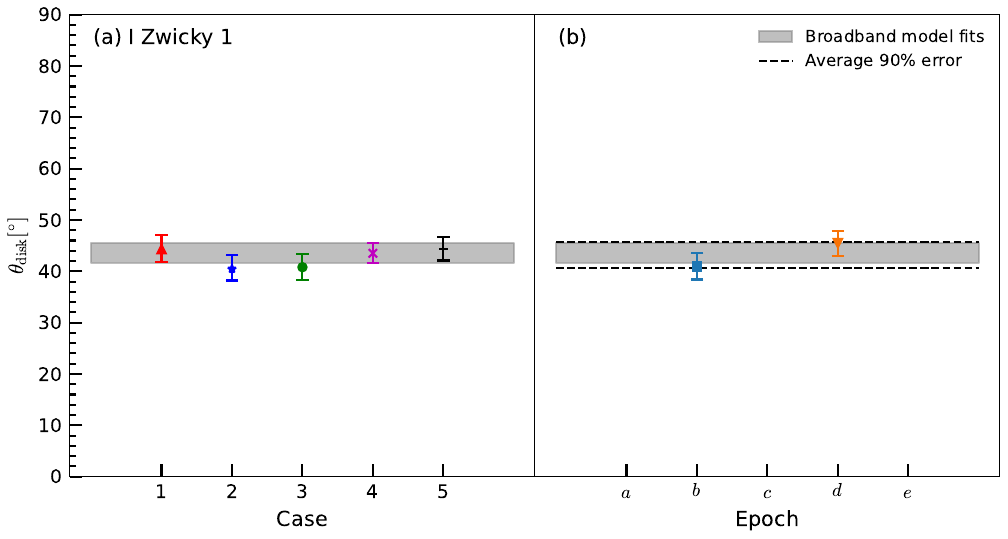}
\caption{A scatter plot showing inclination measurements from (a) different combinations of I\,Zwicky\,1 data groups (cases~1--5) and (b) two simultaneous epochs ($b$ and $d$). The areas shaded in gray show the confidence range for $\theta_\mathrm{disk}$ derived from broadband model fitting (Section~\ref{sec:modelfit}). The two dashed lines give the average of the upper and lower edges of the 90\% error bars of the two epochs.}
\label{fig:izw_comb}
\end{figure*}

\begin{figure*}
\centering
\includegraphics{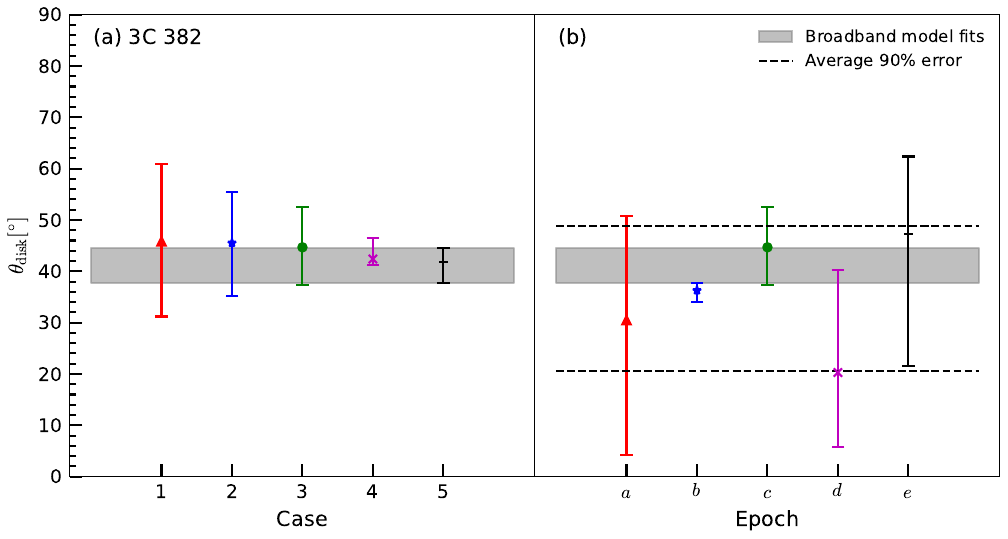}
\caption{A scatter plot showing inclination measurements from (a) different combinations of 3C\,382 data groups (cases~1--5) and (b) five simultaneous epochs ($a-e$). The areas shaded in gray show the confidence range for $\theta_\mathrm{disk}$ derived from broadband model fitting (Section~\ref{sec:modelfit}). The two dashed lines give the average of the upper and lower edges of the 90\% error bars of the five epochs.}
\label{fig:3c_comb}
\end{figure*}

\subsection{Inclination with Different Combinations}
\label{sec:different_comb}
We rearrange the observed data to mimic the different circumstances (cases) discussed in Section~\ref{sec:systematical_method}. For I\,Zwicky\,1, the result from epoch~$b$ is selected to present a case~3 fit. We produce a case~1 fit with the XMM-Newton observation from epoch~$b$, a case~2 fit with the XMM-Newton spectra from epoch~$b$ and NuSTAR spectra from epoch~$d$, and a case~5 fit with epochs~$b$ and $d$. The broadband model fitting in the previous section is a case~4 fit. For 3C\,382, the result from epoch~$c$ is selected to represent a case~3 fit. We conduct a case~1 fit with the XMM-Newton observation from epoch~$c$, a case~2 fit with the XMM-Newton spectra from epoch~$c$ and NuSTAR spectra from epoch~$b$, and a case~4 fit with epochs~$b$ and $c$ and NuSTAR data from epochs~$a$, $d$, and $e$. The broadband model fitting in the previous section is a case~5 fit. Additionally, we produce case~3 fits for all available simultaneous epochs of both sources. Errors are calculated with MCMC, and chain convergence is assured. We show the inclination measurements of the different data combinations for I\,Zwicky\,1 in Figure~\ref{fig:izw_comb}, while the results for 3C\,382 are given in Figure~\ref{fig:3c_comb}. 

The inclination measurements of each source are internally consistent, in view of the overlap of the confidence intervals. The consistency of measurements for all the different cases and different epochs confirms the overall reliability of our method. Comparing the results from different cases, it is clear that the more abundant the data, the smaller the errors, which means that more information provided by the broadband spectra does lead to a better fit, hence a more precise measurement. Moreover, if the several groups of spectra are simultaneous, the dimension of the model is reduced, which further diminishes the flexibility and complexity of the model and improves the fits. Examining the results from different epochs reveals that the measured value from the broadband model fitting in Section~\ref{sec:modelfit} lies within the average value of the five simultaneous epochs. Inclination measurements for both sources can be obtained consistently despite the differences in their outflow structure. Yet, the accuracy of the measurement can be improved if the reflection features in the source spectra are more dominant. For I\,Zwicky\,1 ($R_f > 1$), the inclination measurements from different cases and different epochs only vary a little with confidence intervals mostly overlapping, especially for the two simultaneous epochs, whereas the inclination measurements of 3C\,382 ($R_f \lesssim 0.5$) fluctuate between different combinations of data. We note that combining more data improves the fits more significantly for 3C\,382 than for I\,Zwicky\,1. That said, although case~5 is intuitively the most ideal one among the five cases, the na\"{i}ve expectation that more data give better performance does not necessarily hold. We have shown that applying \textsc{RELXILL} to single-epoch data that exhibit strong reflection signal (I\,Zwicky\,1) suffices to offer satisfactory results with errors of only several degrees. Yet, for data with weak reflection features like those in the single epochs of 3C\,382, the fits can yield dubious results with measured inclination angles varying by tens of degrees. 

In summary: inner accretion disk inclinations can be measured more consistently if the reflection component is more dominant in the spectra or if more observations are combined and fitted simultaneously.

\section{Fitting Simulated Spectra}
\label{sec:simulate}

\subsection{Simulator Settings}
\label{sec:sim_set}
We produce mock spectra with the model \textsc{constant*(zBBody+RelxillCp)} using the \texttt{fakeit} method in \textsc{XSPEC}, as implemented in the Python interface \textsc{PyXspec v.2.1.0} \citep{arnaud2016}. To mimic simultaneous, single-epoch observations (case~3) of nearby AGNs, each group of spectra is generated in the $0.3-10\,\rm keV$ band of XMM-Newton/EPIC and the $3-78\,\rm keV$ band of NuSTAR, using response files from the epoch~$c$ observations of 3C\,382 and their respective ancillary files. The source exposure and background exposure time are all 25\,ks. As the background flux of the simulated NuSTAR spectra overwhelms the source flux at $\sim 50\,\rm keV$, we ignore energy bins above $50\,\rm keV$ for the mock FPMA/B spectra. The mock spectra are generated by setting $R_\mathrm{in}$ to the radius of the innermost stable circular orbit, $R_\mathrm{out} = 400r_{g}$, and $n = 10^{15}\,\mathrm{cm^{-3}}$. The blackbody temperature is set to 0.1\,keV, the redshift to $z=0.05557$ \citep{bosch2015}, and other parameters are initialized as shown in Table~\ref{tab:sim_grid}. 

We study all combinations in Table~\ref{tab:sim_grid}, creating 13,860 distinct simulations. The range for $\theta_\mathrm{disk}$ covers nearly face-on to nearly edge-on systems, which have a high degree of relativistic smearing. The input values of the iron abundance $A_\mathrm{Fe}$ are solar to super-solar to ensure prominent Fe features and to represent the physical conditions in most observations \citep{garcia2013}. The photon index $\Gamma$ and ionization $\xi$ lie within the physical range in AGNs. The thermal temperature $kT_{e}$ of the coronal electrons shapes the high-energy cutoff. Although the physical nature of the corona currently cannot be derived from first principles (for progress in magnetohydrodynamics simulations, see \citealt{kinch2016,kinch2020} and \citealt{jiang2019}), a firm upper limit on $kT_{e}$ is given by the lower limit of $\sim$511\,keV imposed by the temperature of electron-positron pair production. Observations typically find coronal temperatures of $\sim 150\,\rm keV$ \citep[e.g.,][]{ricci2017}, and the cutoff energy set in several models is as high as $300\,\rm keV$ to even $\sim 500\,\rm keV$ \citep[e.g.,][]{balokovic2020,kamraj2022}. 

The radial emission profile, which is degenerate with $\theta_\mathrm{disk}$, is intricately linked to the selection of coronal models---a topic we delve into in Section~\ref{sec:relselect}. While our primary focus for the mock tests does not involve a comprehensive exploration of the impact of radial emissivity parameters (Index$_\mathrm{1}$, Index$_\mathrm{2}$, $R_\mathrm{br}$) on inclination measurements, we assign them reasonable values and conduct tests to evaluate their successful recoveries.

The spin parameter influences the entire simulation grid. We first select a specific value of $a_\ast$ for preliminary discussions on other parameters, and subsequently extend the coverage from $a_\ast = -0.998$ to $+0.998$. Theoretically, higher values of $a_\ast$ usually allow for a larger grid for $R_f$ and $\theta_\mathrm{disk}$ (see Section~\ref{sec:spin}). In the reflection modeling of actual observations, reported spin values often lean toward higher values \citep[$a_\ast > 0.9$;][]{reynolds2021}. Therefore, exploring the parameter space with a higher spin is more informative. Moreover, the degeneracies intertwined within $a_\ast$ that affect the retrieval of other parameters are fortunately disentangled with higher spin, as accurate retrieval of the spin is typically achieved for values above $0.8$ in mock analyses \citep{bonson2016,choudhury2017,kammoun2018}. Thus, we initially set $a_\ast$ to its theoretical maximum of 0.998 to mitigate the influence of spin degeneracy and study this more flexible slice of the parameter space. In Section~\ref{sec:spin}, we extend the discussion to other values of $a_\ast$ (Table~\ref{tab:sim_grid}) to assess the effect of spin degeneracy on the precision of inclination measurement.

\begin{deluxetable}{cc}[t]
\tablenum{9}
\setcounter{table}{9}
\caption{Parameter Values of the Simulation Grid}
\label{tab:sim_grid}
\tablehead{
\colhead{Parameter} &
\colhead{Input Range}
}
\startdata
$\theta_\mathrm{disk}$ & [3, 15, 30, 45, 60, 75, 87]$^\circ$\\
$A_\mathrm{Fe}$ & [1, 3, 5, 7, 9] $A_\odot$\\
$\Gamma$ & 2\\
$\log \xi$ & 3\\
$kT_{e}$ & [30, 150, 300]\,keV\\
Index$_\mathrm{1}$ & 7\\
Index$_\mathrm{2}$ & 3\\
$R_\mathrm{br}$ & 10 $r_{g}$\\
$a_\ast$ & [$\pm0.998$, $\pm0.9$, $\pm0.7$, $\pm0.5$, $\pm0.3$, $0$] \\
$R_f$ & [0, 0.1, 0.3, 0.5, 0.7, 0.9, 1, 3, 5, 7, 9, 10]
\enddata
\end{deluxetable}

\begin{figure}
\centering
\includegraphics[width=\linewidth]{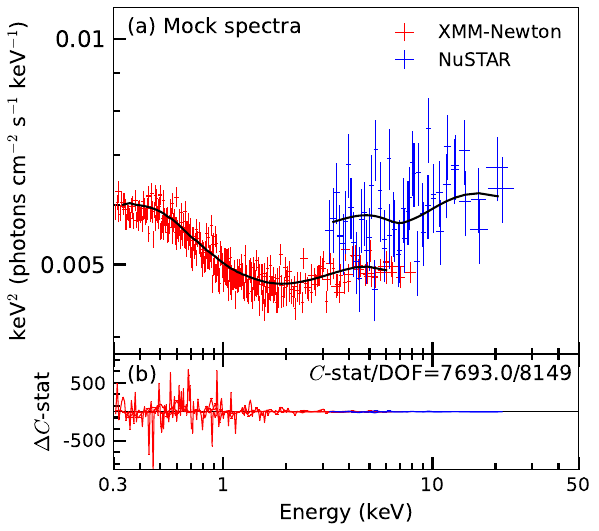}
\caption{(a) An example of the simulated case~3 spectra, with the model in black lines. (b) The $C$-stat residuals between the generated spectra and the model. The input values are $\theta_\mathrm{disk} = 45^\circ$, $a_\ast = 0.998$, $A_\mathrm{Fe} = 3$, $R_f = 1$, and $kT_{e} = \rm 150\,keV$. For plotting purposes, the XMM-Newton data have been rebinned to $\rm S/N > 30$, and the NuSTAR data have been rebinned to $\rm S/N > 10$.}
\label{fig:mock_eg}
\end{figure}

\begin{figure*}
\centering
\includegraphics[width=\linewidth]{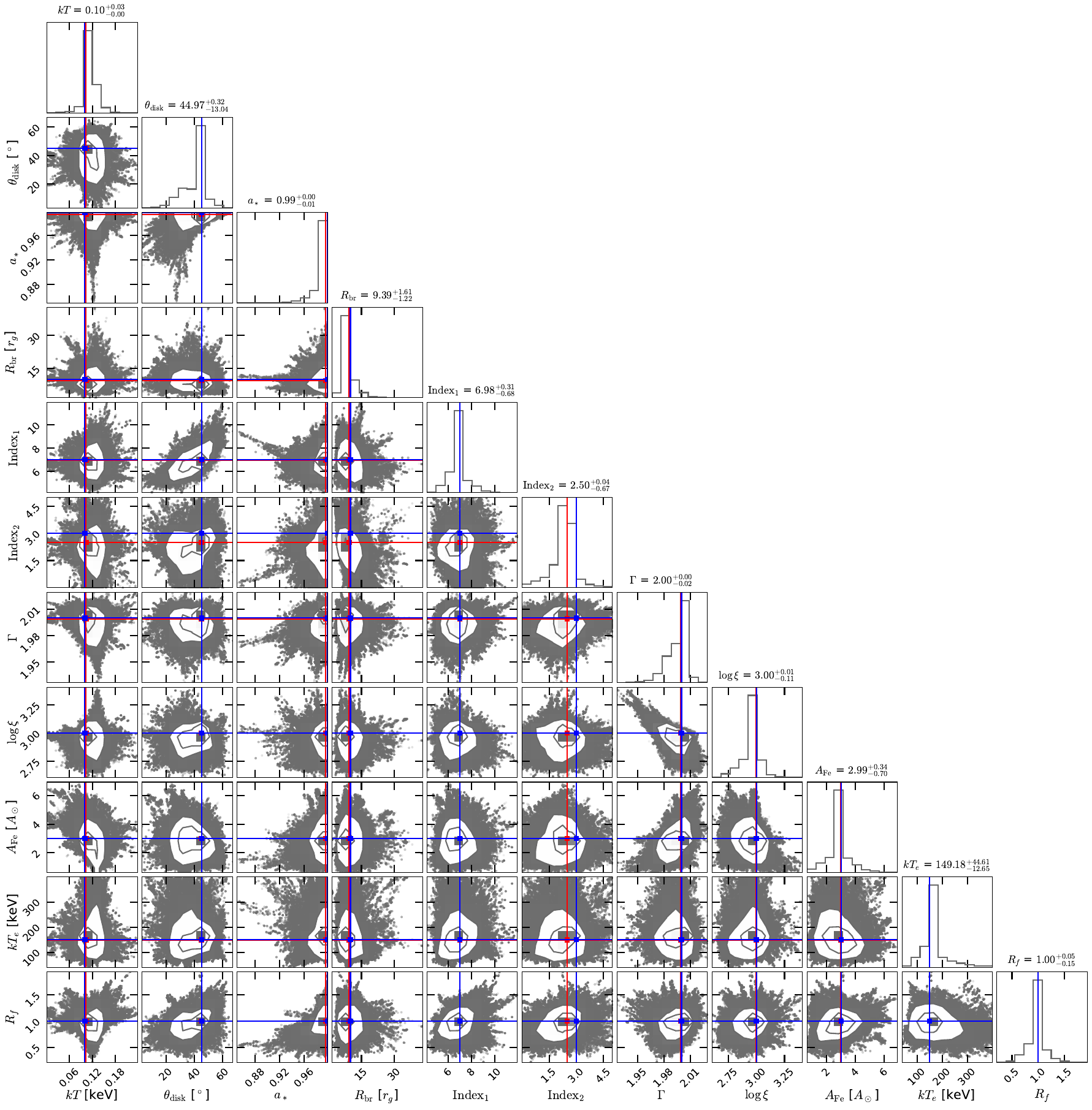}
\caption{Results of the MCMC analysis for the relevant parameters of the spectra shown in Figure~\ref{fig:mock_eg}. The red lines correspond to the measured values, and the blue lines give the input values of the simulation. In general, all parameters are well recovered.}
\label{fig:mock_fit_eg}
\end{figure*}

\begin{figure}[t]
\centering
\includegraphics[width=\linewidth]{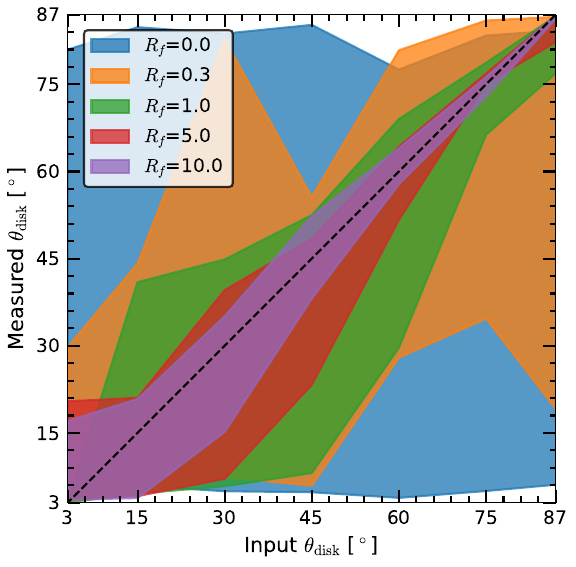}
\caption{Inclination measurements for simulated spectra with solar iron abundance ($A_\mathrm{Fe}=1$) and  mild coronae ($kT_{e} = 30\,\mathrm{keV}$) in the maximum Kerr case ($a_\ast = 0.998$). The colored regions depict the 90\% confidence intervals for different input reflection fractions $R_f$. The dashed line depicts the expected values of $\theta_\mathrm{disk}$.}
\label{fig:simulfit_r}
\end{figure}

\subsection{Mock Analysis}
\label{sec:mock_resu}
Retrieving results from mock spectra can be challenging. Since we know the input parameters in advance, it is technically equivalent to the circumstance that we have already achieved best-fit values when fitting the observed spectra. Nevertheless, knowing the inputs is an idealized situation, for in practice we have little confidence that best-fit parameters are actual model parameters (global minima), although prior knowledge of the physical properties of the source may serve as a guide. Hence, we need to calculate properly the posterior distribution in mock analysis, such that the sampler explores the parameter space as much as possible. However, for our model with much flexibility and complexity, the sampler can get stuck easily in local minima if we start at a random point very far away from the actual values \citep[e.g.,][]{bonson2016}. Therefore, it is problematic to initialize the sampler at random values or to start with the known model values.

We analyze the spectra under technical settings of case~3, where basically all parameters in Table~\ref{tab:sim_grid} and the blackbody temperature are free. As in Section~\ref{sec:modelfit}, errors on the parameters are calculated using the Goodman-Weare MCMC algorithm in \textsc{XSPEC}. Following \citet{choudhury2017}, we set a Gaussian proposal distribution initialized at $\mathbf{X} + 3 \mathbf{\sigma_X}$, where $\mathbf{X}$ is the input model parameter and $\mathbf{\sigma_X}$ is the standard deviation estimated from the covariance matrix of a preliminary fit. For boundary inputs $\theta_\mathrm{disk}=[3^\circ, 87^\circ]$ and $R_f=[0,10]$, the covariance information is changed to a diagonal matrix constructed from 100 times the step length of the parameter. For each group of the mock spectra, we use 200 walkers for the chain and sample to a total of $1.2 \times 10^6$ elements (6000 iterations) with the first $1.2 \times 10^5$ rejected. As an example of the mock and error calculation, we show a group of spectra in Figure~\ref{fig:mock_eg} and its corresponding MCMC corner in Figure~\ref{fig:mock_fit_eg}.

We check the parameter recovery of the mock spectra by seeing whether the ``measured'' (fitted) values are close to the input values, and whether the error range includes the input. Figure~\ref{fig:simulfit_r} compares the measured versus input values of inclination, with colored regions depicting 90\% confidence intervals: the dispersion suggests that the input values are generally recoverable. There is little dependence on the iron abundance or coronal temperature. Since our simulation grid is five-dimensional, the figures in the following sections are provided to show the trend of parameter recovery regardless of the specific parameter choice in one or two dimensions. Additionally, larger reflection fraction and inclination both lead to tighter parameter constraints, as to be expected because either large $R_f$ or large $\theta_\mathrm{disk}$ increases the relativistic smearing of the intrinsic spectra. \cite{dauser2014} find a similar trend for the reflection model based on the lamp-post geometry \citep{matt1991,martocchia1996,reynolds1997,miniutti2004}. The confidence interval shrinks most markedly when $R_f$ varies from 0.3 to 1, with less dramatic but still notable improvements for $R_f = 5-10$. Not surprisingly, the completely reflection-free case ($R_f = 0$) can barely measure $\theta_\mathrm{disk}$; the posterior distribution (blue region) is very close to a uniform distribution. 

\begin{figure*}[t]
\centering
\includegraphics[width=\linewidth]{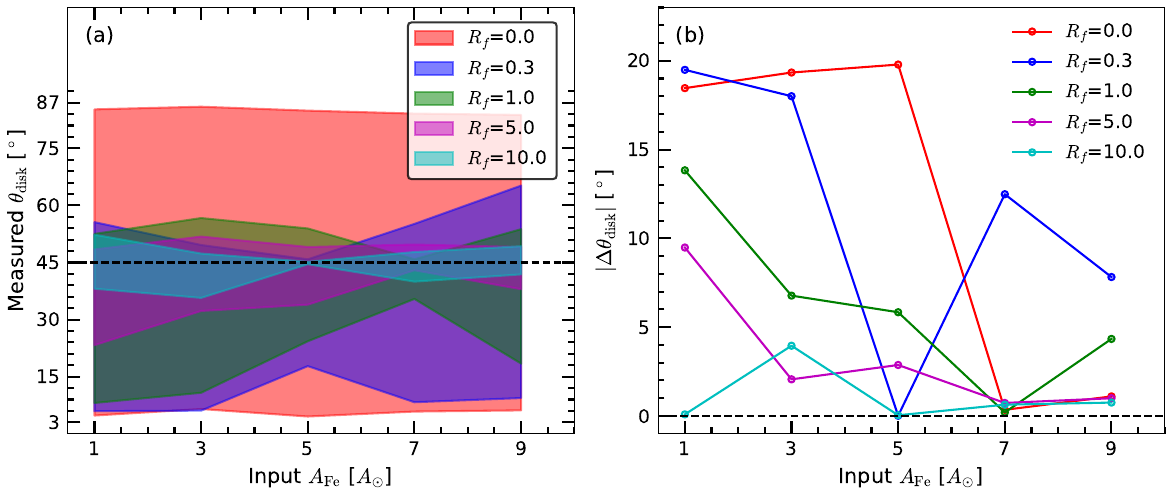}
\caption{The dependence of inner disk inclination $\theta_\mathrm{disk}$ on input iron abundance $A_e$ and reflection fraction $R_f$, for fixed $kT_e = 30 \rm \,keV$, $\theta_\mathrm{disk} = 45^\circ$, and $a_\ast = 0.998$. Panel (a) shows the measurements of $\theta_\mathrm{disk}$ and the 90\% confidence intervals, and panel (b) the absolute offset in $\theta_\mathrm{disk}$.}
\label{fig:afe_error}
\end{figure*}

\begin{figure*}
\centering
\includegraphics[width=\linewidth]{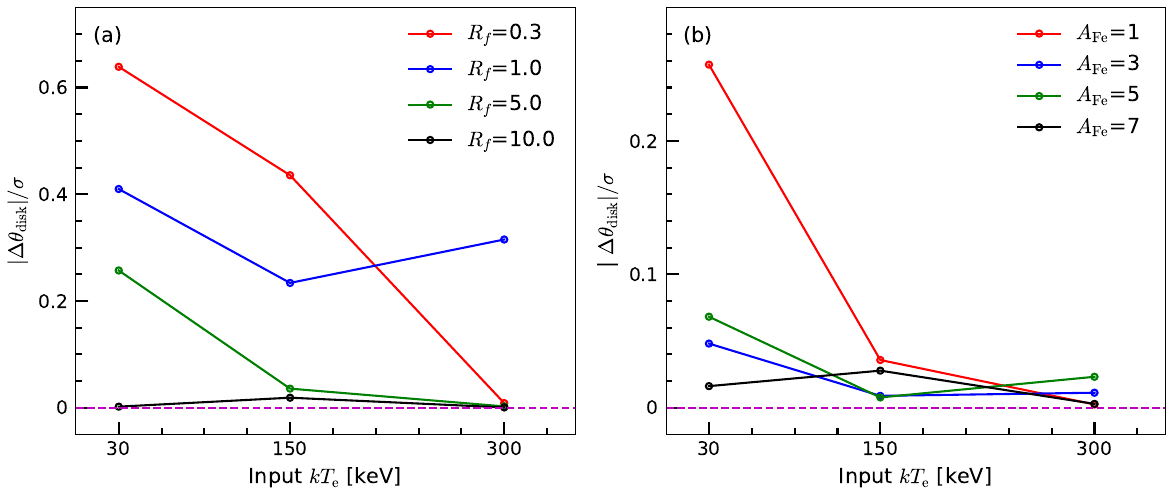}
\caption{The dependence of the relative offset in inner disk inclination ($\sigma$ refers to the average statistical error for each point in each panel) on coronal temperature $kT_e$ and input (a) reflection fraction $R_f$ (with solar $A_\mathrm{Fe}$) and (b) iron abundance $A_e$ (with $R_f = 5$), for fixed $\theta_\mathrm{disk} = 45^\circ$ and $a_\ast = 0.998$.}
\label{fig:kte_error}
\end{figure*}

\section{Discussion}
\label{sec:discussion}

\subsection{Sensitivity to Reflection Fraction}
\label{sec:rf}
Envisioning the corona as a razor-thin layer that hugs the entire disk, the \textsc{RelxillCp} model defines the reflection fraction as
\begin{equation}
R_f \coloneqq \frac{f_\mathrm{disk}}{f_\mathrm{\infty}}, 
\end{equation}\noindent
where $f_\mathrm{disk}$ denotes the fraction of photons hitting the disk and $f_\mathrm{\infty}$ the fraction of photons escaping to infinity \citep{dauser2016} in the rest-frame of the corona. Given that the corona precisely overlays the disk with a large covering fraction, we anticipate the reflection fraction to be $\sim 1$. Therefore, in our spectral fits, $R_f$ is not intended to serve as an indicator of the geometry or relativistic light-bending; instead, it serves as an empirical representation of the relative strength of the reflection component, encompassing the soft excess, the Fe~K$\alpha$ feature, and the Compton hump. Although this parameter lacks physical significance, it is nevertheless important because it provides insight into the proportion of the reflected emission ($\frac{R_f}{1 + R_f}$). This, in turn, helps us to estimate the systematic error and lends confidence in applying the reflection model.

Referring back to the mock results, in a typical case of $R_f=1$ (e.g., Figure~\ref{fig:simulfit_r}), the relative error can be as large as 50\% to 90\% when $\theta_\mathrm{disk}$ decreases to $\lesssim 60^\circ$. However, for larger $\theta_\mathrm{disk}$, the error is $\lesssim 10^\circ$. When $R_f \gtrsim 5$, the proportion of reflection is already larger than 83\%, and further increasing $R_f$ does not significantly raise this proportion, resulting in negligible changes to the confidence interval. On the other hand, for $R_f \lesssim 0.3$, where the reflection signal is weak ($\lesssim 23\%$), robust inclination measurements are challenging to achieve. The impact of reflection fraction is also seen in Section~\ref{sec:modelfit}, where the results of $\theta_\mathrm{disk}$ from different simultaneous epochs of I\,Zwicky\,1 are more consistent compared to 3C\,382. Indeed, even a single-epoch XMM-Newton observation of I\,Zwicky\,1 offers a fair constraint on inclination. Without loss of generality, we define a spectrum as reflection-dominant when $R_f \gtrsim 5$ and reflection-subordinate when $R_f \lesssim 0.3$. In general, it is usually easier to achieve tighter constraints on inner disk inclination for sources with higher $R_f$, and, of course, larger inclination, because the observed spectrum of more edge-on systems naturally has a stronger reflection signal. However, it is also worth noting that combining more groups of data can counterbalance the poor constraints of reflection-subordinate sources, as examplified by the case of 3C\,382. 

\subsection{Sensitivity to Iron Abundance}
\label{sec:afe}
Figure~\ref{fig:afe_error} examines the effect of iron abundance on the measurement of inner disk inclination. Large $A_\mathrm{Fe}$ increases the strength of the whole Fe profile \citep{garcia2013}. In our simulations, both the absolute value of inclination offset and the confidence interval of $\theta_\mathrm{disk}$ generally decrease with a larger input value of $A_\mathrm{Fe}$. Consistent with the results of previous sections, this trend holds as well for smaller $R_f$, although the constraints on $\theta_\mathrm{disk}$ are significantly looser because reflection spectroscopy is intrinsically less reliable for reflection-subordinate cases. In contrast with $R_f$, the influence of $A_\mathrm{Fe}$ on the accuracy of spectrum fitting is not as significant, as evidenced by comparing Figure~\ref{fig:simulfit_r} and Figure~\ref{fig:afe_error}b, or by comparing the effect of $R_f$ and $A_\mathrm{Fe}$ in Figure~\ref{fig:afe_error} itself. On the one hand, the inclination results are rather reliable already for sources with solar or mildly super-solar abundance; on the other hand, it implies that the model is not that sensitive to $A_\mathrm{Fe}$ compared to $R_f$. This, aside from the support for super-solar abundance in AGN reflection modeling \citep{fabian2006}, also explains why the broadband fitting for 3C\,382 returns such a huge value of $A_\mathrm{Fe}$ with a large error even though we simultaneously fit five epochs. We cannot use $A_\mathrm{Fe}$ as a basis for judging the robustness of reflection models.

\subsection{Sensitivity to Coronal Temperature}
\label{sec:kte}
Figure~\ref{fig:kte_error} documents the effect of coronal temperature on the measurement of inner disk inclination. The inclination offset in terms of error generally decreases with larger input values of $kT_{e}$. Consistent with previous sections, this trend holds regardless of the values of $R_f$ and $A_\mathrm{Fe}$, although the effect from $R_f$ is greater than that from $A_\mathrm{Fe}$. Similar to the case of $A_\mathrm{Fe}$, the inclinations are fairly reliable for sources with $kT_e = 30\, \rm keV$. In all cases, however, $|\theta_\mathrm{disk}|/\sigma < 1$, meaning that deviations between the measured and input inclinations all lie within the $1\sigma$ confidence.

\begin{figure*}[t]
\centering
\includegraphics[width=\linewidth]{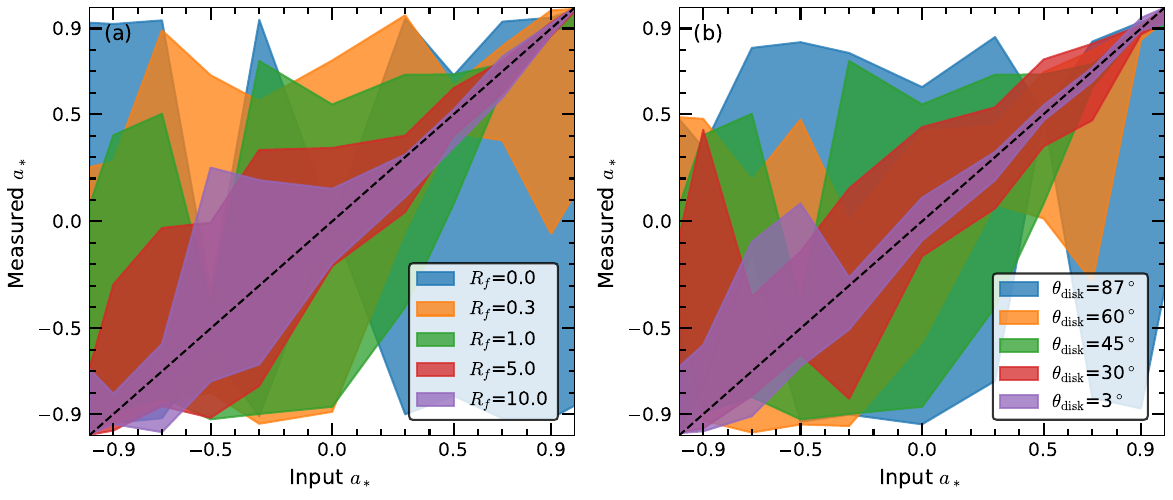}
\caption{Spin measurements for simulated spectra with different values of (a) reflection fraction $R_f$ and fixed $\theta_\mathrm{disk} = 45^\circ$, and (b) different values of $\theta_\mathrm{disk}$ and fixed $R_f=1$. In all panels, we fix $A_\mathrm{Fe}=1$ and $kT_{e} = 30\, \rm keV$.}
\label{fig:a_1}
\end{figure*}
\begin{figure*}
\centering
\includegraphics[width=\linewidth]{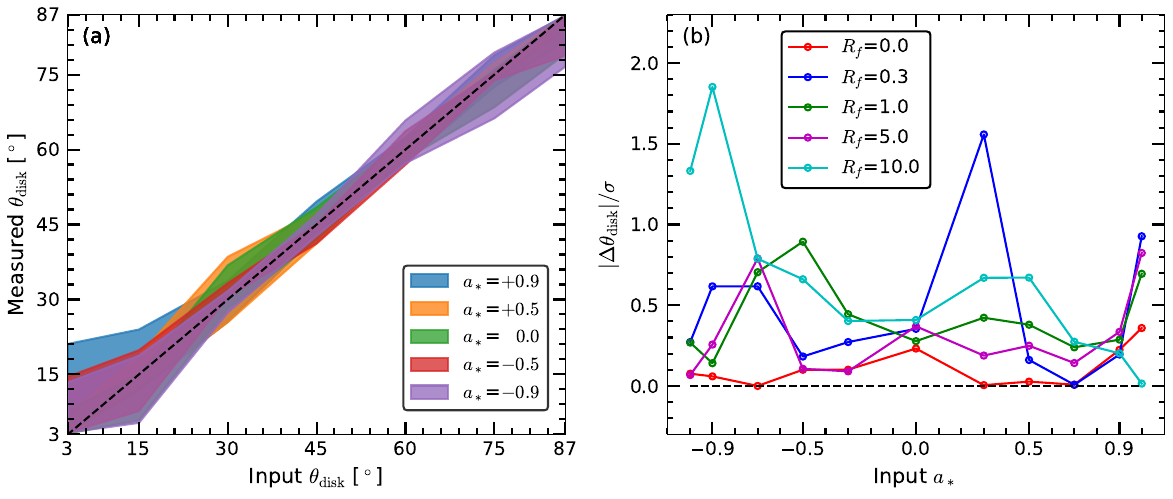}
\caption{Panel (a) shows the dependence of the measurements of $\theta_\mathrm{disk}$ on spin $a_\ast$ with $R_f=1$; the colored regions depict the 90\% confidence intervals and the dashed line the expected values. Panel (b) illustrates the relative offset in $\theta_\mathrm{disk}$ with respect to different $a_\ast$ and $R_f$. In all panels, we fix $A_\mathrm{Fe}=1$ and $kT_{e} = 30\, \rm keV$.}
\label{fig:a_2}
\end{figure*}

\subsection{Spin Degeneracy}
\label{sec:spin}

Both the spin and the inner disk inclination contribute to the broadening of the Fe~K$\alpha$ emission line, with the red edge of the profile being more sensitive to $a_\ast$ and the blue edge more sensitive to $\theta_\mathrm{disk}$. Different combinations of $a_\ast$ and $\theta_\mathrm{disk}$ can yield similar Fe profiles, making them challenging to distinguish during spectral fitting. This inherent degeneracy is further complicated by the coupling of both parameters with the entire model. Given the intricate interplay between $a_\ast$, $\theta_\mathrm{disk}$, and other model parameters, our mock tests are conducted with specific considerations for two corona-related parameters ($\xi$ and $\Gamma$), disk geometry, and, once again, the physical background of $R_f$:

\begin{enumerate}
\item While $\xi$ is better measured with higher values, incorrect measurements of $\xi$, followed by $A_\mathrm{Fe}$, $\Gamma$, and $\theta_\mathrm{disk}$, pose challenges for spin retrieval \citep{choudhury2017}. We use $\log \xi = 3$ as the input and find that the 90\% error of $\xi$ is $\lesssim 300\, \rm erg~cm~s^{-1}$ for all mock spectra.

\item According to \citet{choudhury2017}, underestimating $\xi$ corresponds to overestimating $\Gamma$, and vice versa. Lower $\xi$ shifts the spectra toward harder X-rays, as larger $\Gamma$ produces less soft X-ray emission. Here, $\xi$ influences recombination and photoionization strength, directly shaping the recombination continuum. We generate spectra with $\xi = 1000 \, \rm erg~cm~s^{-1}$ as an input value that can be acceptably retrieved, reducing its influence on $a_\ast$ and the overall model fitting. As the power-law exponent in the spectral model, $\Gamma$ should be easier to independently constrain in our broadband fits.

\item The value of $a_\ast$ sets a physical lower limit for the inner radius of the accretion disk with a negative correlation, which is the main effect of the spin in observations \citep{dauser2010}. Besides, changing $a_\ast$ modifies the geodesics in ray tracing, which has a more complicated influence on observed spectra. For a disk geometry, $a_\ast=0.998$ supports a disk reaching $\sim 1.24 \,r_g$, while $a_\ast=-0.998$ pushes the innermost stable circular orbit to $9\,r_g$. We set the broken radius to $10 \,r_g$ to avoid physically impossible circumstances of $R_\mathrm{br}<R_\mathrm{in}$.

\item The $R_\mathrm{in}-a_\ast$ relation also affects the reflection fraction. The maximal possible $R_f$ declines for smaller $a_\ast$ or larger $R_\mathrm{in}$ \citep{dauser2014}. Although our phenomenological $R_f$ is not directly constrained by $a_\ast$, some extreme combinations in our simulation grid are rare in real observations.
    
\item Aside from the spin, measured $R_f$ is further affected by $\theta_\mathrm{disk}$ because the observer only receives flux traveling at a certain orientation with respect to the system. According to \citet{suebsuwong2006}, extremely large $R_f$ ($\gtrsim 10$) is possible only for systems with high inclinations ($\theta_\mathrm{disk} \gtrsim 70^\circ$). Still, since $R_f$ in \textsc{RelxillCp} is not physical, a few extreme combinations are possible in mocks but do not necessarily exist. The only motivation for including those points in the parameter space is to make the discussion more complete computationally.
\end{enumerate}
\noindent
In conclusion, we provide rationale for selecting the specific values for $\xi$, $\Gamma$, and $R_\mathrm{br}$ summarized in Table~\ref{tab:sim_grid}. Additionally, our previous analyses indicate that $A_\mathrm{Fe}$ and $kT_e$ have a lesser impact on $\theta_\mathrm{disk}$ derivation compared to $R_f$. Consequently, the ensuing discussions will focus on the interplay between spin and inclination measurements and the influence of the reflection fraction. 

Like the case for inclination, the measurement of $a_\ast$ is influenced by its intrinsic value. All three preceding studies on mock spectra of \textsc{RELXILL} focused on the reliability of spin measurements and observed a more accurate recovery of $a_\ast$ with higher $R_f$ and larger $a_\ast$. This general trend is confirmed in our analysis, as depicted Figure~\ref{fig:a_1}a. A larger $a_\ast$ allows the inner edge of the disk to approach the BH more closely, resulting in more pronounced redshifting of photons and the formation of a broader Fe line that is easier to detect. Figure~\ref{fig:a_1}b examines the effect of inclination on spin measurement, with colored regions depicting 90\% confidence intervals. Generally, larger inclinations lead to less constrained spins because the differences in line broadening caused by $a_\ast$ are more evident at smaller $\theta_\mathrm{disk}$ \citep[e.g., see][]{dauser2010}. The inclination measurements are affected minimally by the spin (Figures~\ref{fig:a_2}a and \ref{fig:a_2}b). 

Considering the entire simulation grid of 13,860 spectra, 85\% exhibit an offset between the input and measured inclination smaller than $10^\circ$, 77\% have an inclination offset smaller than $5^\circ$, and 62\% have an absolute inclination offset smaller than $5^\circ$, with their 90\% confidence interval covering the input inclination. All reduced fit statistics are below 1.1.

\subsection{Comparison with Previous Work}
\label{sec:comparison}
In this section, we compare both observed and simulated spectra analyses with previous work. Overall, eight X-ray studies have been reported for I\,Zwicky\,1 since the advent of XMM-Newton. For 3C\,382, four works have been reported with XMM-Newton observations. Three mock analyses have been published for the \textsc{RELXILL} family, all focusing on the recovery of $a_\ast$.

For I\,Zwicky\,1, \citet{gallo2004} discovered a broad iron line, a hard X-ray flare, and spectral hardening during the flare with the $\sim 20\, \rm ks$ XMM-Newton observation taken in 2002. \citet{costantini2007} discovered two WAs in soft X-rays from the RGS data, which was further confirmed by \citet{silva2018} from their two XMM-Newton observations in 2015. Based on the same observations in our work, \citet{wilkins2021} and \citet{wilkins2022} suggested that both the broad Fe~K$\alpha$ line extending to 3\,keV and the Compton hump around 25\,keV were well explained with the disk reflection model. Using a lamp-post coronal geometry, an ionization gradient, and a twice-broken power-law emissivity, they obtained a corona height of $h > 14 \,r_g$ during the flares and $h\approx 3.7 \,r_g$ before and after flares. \citet{ding2022} continued with the idea of reflection and showed that a single broken power-law emissivity implemented by \textsc{RelxillCp} was sufficient, and that an ionization gradient was unnecessary once the absorbers were added correctly (see their Figure~2 and Section~3.2). While most of the reflection parameters ($\Gamma$, $kT_e$, $R_f$) indicated the same evolutionary trend as in \citet{wilkins2022}, their result for the spin ($a_\ast>0.973$) is consistent with that ($a_\ast>0.94$) of \citet{wilkins2022}; their inclination angle $\theta_\mathrm{disk} \approx 42.3^\circ$ is $\sim 10^\circ$ smaller; and their $R_f \gtrsim 1$ is an order of magnitude bigger, which was argued to be the consequence of applying a different ray-tracing kernel. Modeling the reflection continuum and ionized absorbers as in \citet{ding2022} yields generally similar results. However, we provide measurements for Index$_2$ ($\sim 3$) while they fixed it to 3. Including ultraviolet data from XMM-Newton observations, \citet{rogantini2022} fit the spectral energy distribution via both a frequentist and a Bayesian method and further acknowledged the components found in previous works. With \texttt{refl} in \textsc{SPEX} \citep{kaastra1996} assuming solar metal abundance and zero ionization, they reported a reflection scale factor (equivalent to $R_f$) of $0.43$. 

\citet{reeves2019} analyzed the 2015 XMM-Newton observations (3--10\,keV) with four different models consecutively: a P~Cygni profile \citep{done2007}, a photoionization continuum, a reflection model (\textsc{XILLVER}$\ast$\textsc{KDBLUR}), and a fast ($\sim c$), wide-angle disk wind model \citep{sim2008}. They claimed that the broad Fe~K profile and the deep 9\,keV absorption can be described by a disk wind, although the reflection component could not be excluded. Their reflection modeling returned an ionization $\xi \approx 10^3$, a reflection fraction $\sim 0.6$, a flat single power-law emissivity with an index around 2.2, and an inclination angle even 2--20 degrees larger than that of \citet{wilkins2022}. However, their spectra excluded the hard X-rays, and their model did not have spin as a parameter due to the use of another ray-tracing kernel. On modeling the outflowing wind, they set the wind location to the escape radius of the BH and obtained an accretion disk inclination angle $49.7^\circ$ as an indicator of blueshifted absorption features.

For 3C\,382, \citet{torresi2010} resolved a WA in the RGS spectra from the 2008 observation, which was the first observed WA in a broad-line radio galaxy. \citet{sambruna2011} analyzed the EPIC spectrum from the same observation using parameters derived from spectral fits of joint Suzaku and Swift/BAT data and discovered both a broad and narrow Fe~K$\alpha$ line, a hump above 10\,keV, a soft excess, and a WA. They argued for a two-ionization reflection with a highly ionized component reproducing the broad emission line and a mildly ionized component accounting for the narrow emission line, the hump, and the soft excess. Both reflection components contributed to $\sim 10\%$ of the total continuum. Using two \textsc{XILLVER} models, they reported $\log \xi = 1.54$ and $2.93$. Due to low S/N, they could not constrain the spin with both \textsc{RELCONV} and \textsc{RELLINE}. However, for an assumed emissivity index of 2.5, the inner and outer radii of the reflection region were around $R_\mathrm{in} \approx 10 \,r_g$ and $r_\mathrm{out} \approx 25 \,r_g$, respectively, and the inner accretion disk inclination is around 30 degrees. With the simultaneous NuSTAR and Swift observations in 2012 and the NuSTAR observation in 2013, \citet{ballantyne2014} found a weak Fe line in the NuSTAR spectra and argued that it originated from the outer regions ($\gtrsim 50 \,r_g$) of an outflowing corona and ionized accretion disk. In their attempts to utilize \textsc{RELXILL}, they reported an upper limit for the reflection fraction ($R_f \lesssim 0.1$) and a lower limit for the inclination angle ($\theta_\mathrm{disk} > 25^\circ$), which was insensitive to the data. Based on the same observations in our work, \citet{ursini2018} jointly fit the PN and NuSTAR spectra and argued that the line broadening can be explained by instrumental effects or reflection on the torus or the broad-line region. With \textsc{RELXILL}, they reported a very small reflection fraction consistent with that estimated by \citet{ballantyne2014}, and yet an upper limit of $15^\circ$ for the inclination angle. They also calculated the intrinsic line width for all five epochs: $\sigma <0.2$, $<0.35$, $0.5^{+0.3}_{-0.2}$, $ 0.13^{+0.09}_{-0.08}$, and $< 0.2$, respectively. 

\citet{bonson2016} generated over 4000 high-quality mock Seyfert~1 spectra to study spin recovery with \textsc{RELXILL}, assuming reflection fractions of 1 and 5. They discovered that most parameters were overestimated with spectral coverage limited to 2.5--10.0\,keV and that extending the data to 70\,keV improved the measurements. However, the model was found insensitive to Index$_1$, and $a_\ast$ was well constrained only for large values of $a_\ast$ (e.g., with error $\pm 0.10$ for input $a_\ast> 0.8$). \citet{choudhury2017} tested the recovery of $a_\ast$ and $\xi$ in the lamp-post flavor of \textsc{RELXILL}--\textsc{RelxillLp}--with 5800 NuSTAR simulations of a bright source. They found that both parameters are well-recovered at 90\% confidence with improving constraints at higher $R_f$ and $a_\ast$, and lower source height $h$. Besides, they stressed the importance of choosing a reasonable initial point for spectral fitting and spanning the parameter space carefully for nonlinear-behaving parameters (e.g., $\xi$) to avoid the fit being trapped in a local minimum. \citet{kammoun2018} generated 60 mock spectra with more complex components including warm and neutral absorbers, relativistic (\textsc{RelxillLp}) and distant reflection, and thermal emission. Their blind fit was produced starting with selecting models, which emulated the typical procedure of observed spectral fitting. They reported that neither the absorbers nor the input $a_\ast$ affected the fits significantly, although $h$ was important. Compared with our mock analysis, however, both \citet{bonson2016} and \citet{choudhury2017} excluded the soft band. Tests in \citet{kammoun2018}, although more complete in spectral coverage, included a plethora of spectral models beyond our main focus. More importantly, all three works generate mock spectra with a long exposure time to show the upper limitations of model performance with high counts and high-S/N data. By contrast, our analysis tests the model performance with data settings more closely related to real observations. In addition, \citet{choudhury2017} and \citet{kammoun2018} adopted the lamp-post geometry, whereas we do not assume any particular coronal configuration.

\subsection{Model Selection}
\label{sec:relselect}
In practice, selecting a model to interpret the X-ray observations of AGNs is challenging because their typically complicated spectra can be fit by several candidate models. The comparisons in the previous section clearly illustrate the statistical and potential systematic uncertainties introduced during model selection (aspect 3 in Section~\ref{sec:reliability_method}). For I\,Zwicky\,1, the presence of obvious line broadening and a high-energy hump may seem to provide exclusive evidence for the reflection model, but the wind model cannot be fully excluded. In the case of 3C\,382, a weak reflection signal leaves ample room for alternative models, and the results of reflection models are significantly shifted based on data quality and quantity (Section~\ref{sec:different_comb}), as well as the strategy of parameter selection in fitting. Even within the framework of reflection, the choice of the emissivity profile associated with the coronal model also affects the measurement of inclination and other parameters. Therefore, current reflection modeling cannot circumvent case-by-case studies to fit observed spectra. For reflection-subordinate sources, combining multi-epoch broadband spectra and carefully exploring the parameter space becomes crucial, as it can at least minimize the statistical error.

Within the framework of reflection, we choose \textsc{RelxillCp} over \textsc{RELXILL} or \textsc{RelxillD} due to its thermal Comptonization mechanism of the corona. The main difference among the models in the \textsc{RELXILL} family lies in the configuration of the corona, which might influence inclination measurements because it determines the radial emissivity profile of the system. In practice, the radial emissivity profile can be parameterized for various geometries \citep[e.g.,][]{dauser2013}. For \textsc{RelxillCp}, the emissivity is given as $r^{-\mathrm{Index}_1}$ between $R_\mathrm{in}$ and $R_\mathrm{br}$ and $r^{-\mathrm{Index}_2}$ between $R_\mathrm{br}$ and $R_\mathrm{out}$. The fact that these empirical parameterizations often find a high Index$_1$ ($\gtrsim 5$; see also Section~\ref{sec:modelfit}) has motivated the lamp-post geometry--an isotropically irradiating source located in the vicinity and on the rotational axis of the BH--which has been widely employed in recent years to interpret the X-ray spectra of AGNs \citep[e.g.,][]{wilkins2011,dauser2012,miller2015,beuchert2017}. This leads to the advent of the \textsc{RelxillLp} series. 
We stress, however, that the configuration of the corona remains unclear and difficult to distinguish from spectral analysis \citep[e.g.,][]{tortosa2018}.  Further determination of coronal geometry usually requires additional information, for instance from X-ray reverberation mapping \citep[e.g.,][]{reynolds1999,fabian2009,kara2016,caballerogarcia2020} or X-ray spectropolarimetry (e.g., \citealp{tagliacozzo2023}; but see \citealp{ingram2023}). Comparing well-understood, theoretical emissivity profiles to the profile determined from the Fe line, and applying constraints from reverberation lags, can help establish the geometry \citep{wilkins2012}, but this is beyond the scope of this work.

\begin{deluxetable}{cccccccc}[t]
\tablenum{10}
\caption{Comparison of Best-fit Results of I\,Zwicky\,1 Using \textsc{RelxillCp} and \textsc{RelxillLpCp}}
\setcounter{table}{10}
\tablehead{
\colhead{Flavor} &
\colhead{$\theta_\mathrm{disk}$}\,($^\circ$) &
\colhead{$a_\ast$} &
\colhead{$C$-stat} &
\colhead{DOF} &
\colhead{$C_\mathrm{red}$} &
\colhead{$k$} &
\colhead{AIC} \\
\colhead{(1)} &
\colhead{(2)} &
\colhead{(3)} &
\colhead{(4)} &
\colhead{(5)} &
\colhead{(6)} &
\colhead{(7)} &
\colhead{(8)} 
}
\startdata
\textsc{-Cp}& $44.3_{-2.4}^{+2.8}$ & $>0.978$ & 926 & 829 & 1.117 & 64 & 1054\\
\textsc{-LpCp}& $44.1_{-6.4}^{+8.2}$ & $>0.964$ & 1010 & 824 & 1.226 & 69 & 1148
\enddata
\tablecomments{
Col. (1): Flavor of the model used.
Col. (2): Inclination angle of the inner accretion disk.
Col. (3): Dimensionless BH spin.
Col. (4): Cash statistics.
Col. (5): Degree of freedom.
Col. (6): Reduced $C$-stat.
Col. (7): Number of parameters.
Col. (8): Akaike Information Criterion.
}
\label{tab:izw1_lamppost}
\end{deluxetable}

Aside from the arguments in coronal geometry, \textsc{RelxillLpCp} includes another feature compared to \textsc{RelxillCp}: the ionization gradient. Photoionized by the irradiation from the corona, the accretion disk receives a flux as a decreasing function of the radius, leading to a radial distribution of the ionization parameter,
\begin{equation}\label{eq:ionization}
    \xi \coloneqq \frac{4 \pi F(r)}{n}, 
\end{equation}\noindent
where $n$ is the number density of the plasma receiving an ionizing flux $F$. In \textsc{RELXILL}, the parameter $n$ (or $n_\mathrm{H}$) usually denotes the hydrogen number density, following \textsc{XSTAR}. However, the $n$ in Equation~\eqref{eq:ionization} should be the electron number density, equal to 1.2 times the hydrogen density since most H atoms are ionized, and $F$ becomes the net integrated flux in the 1–1000\,Ry energy range \citep[$F_{\rm X}$;][]{tarter1969,garcia2013}. Thus, the ionization parameter in \textsc{RelxillCp} is $\xi = 4 \pi F_{\rm X} / 1.2 n_\mathrm{H}$. \citet{svoboda2012} argued that the radial ionization profile should be considered to derive the very steep emissivity profiles (Index$_1 \geq 7$), which is degenerate with both the spin and the inclination. 

For 3C\,382, it is pointless to debate whether we need to include additional features because reflection modeling suffers from a low reflection fraction. But it is worthwhile to see if the ionization gradient affects the inclination angle for typical sources like I\,Zwicky\,1. We conduct model comparison via the likelihood ratio ($\Lambda$) and the \citet{akaike1974} information criterion (AIC). Because we use $C$-stat for spectral fitting, the likelihood-ratio test is equivalent to the $\chi^2$ hypothesis test \citep{wilks1938,wilks1963},
\begin{equation}
    - 2 \log \Lambda = \Delta C \to \chi^2_{\Delta k}, 
\end{equation}\noindent
where $k$ is the number of model parameters. We apply AIC to compare the ability of models to properly describe the observed spectrum without significant over-fitting and under-fitting. The AIC is defined by
\begin{equation}
    \mathrm{AIC} \coloneqq -2 \log \mathcal{L} + 2k = C + 2k,
\end{equation}\noindent
which not only includes the likelihood function but also punishes the model with more parameters. We calculate AIC for our fit of I\,Zwicky\,1 using \textsc{RelxillCp} as the null model, and we produce alternative model fitting using \textsc{RelxillLpCp} with coronal parameters initialized according to the best-fit values in \citet{wilkins2022}. During the fit, the density of the disk and the ionization gradient are considered as corona-related parameters, meaning that they are free to vary across spectral epochs. The radial density profile is not considered because its effect is incomparable with that of irradiation. We report the inclination, spin, and other values related to statistical tests in Table~\ref{tab:izw1_lamppost}. As the inclination and spin measurements are not significantly affected, the reduced $C$-stat for \textsc{RelxillLpCp} is slightly larger than for \textsc{RelxillCp}. A likelihood-ratio test cannot reject the null model, and the alternative model is disfavored by $\Delta \mathrm{AIC} = +94$. Therefore, the ionization gradient is unnecessary for I\,Zwicky\,1, which is consistent with \citet{ding2022}. In addition, returning radiation is also included in \textsc{RelxillLpCp} \citep{dauser2022}, which is tested jointly. Without a notable effect on $\theta_\mathrm{disk}$, we have observed stronger reflection ($R_f > 0.5$ to $> 4.3$) and a lower coronal height ($h \lesssim 5\,r_g$ to $\lesssim 2\,r_g$) comparing to \citet{wilkins2022} ($R_f < 0.5$; $h > 14\,r_g$ to $\gtrsim 3\,r_g$), even though all parameters evolve in the same trend.

For generality and our purpose of inclination measurements, it is readily acceptable to adopt \textsc{RelxillCp} to use its two emissivity indices (Index$_1$ and Index$_2$) and the broken radius ($R_\mathrm{br}$), which implicitly has taken into consideration various coronal geometries without further assumptions. However, it is far more difficult to interpret than simply applying a lamp-post geometry with which many of the model parameters are directly calculated \citep[e.g., the relation between $R_f$ and $\theta_\mathrm{disk}$ is well-studied in][]{dauser2014}. Moreover, the model is insensitive to the inner emissivity Index$_1$ \citep{bonson2016}, rendering it challenging to be constrained in a blind fit. We suggest assigning proper initial values for the emissivity parameters and always providing the fitted emissivity along with the inclination measurements for reproducibility (as in Section~\ref{sec:modelfit}).

\begin{deluxetable}{cccccccc}
\tablenum{11}
\caption{Comparison of Best-fit Results of 3C\,382 with Different Treatments of the Soft Band}
\setcounter{table}{11}
\tablehead{
\colhead{Fit} &
\colhead{$\theta_\mathrm{disk}$}\,($^\circ$) &
\colhead{$a_\ast$} &
\colhead{$C$-stat} &
\colhead{DOF} &
\colhead{$C_\mathrm{red}$} &
\colhead{$k$} &
\colhead{AIC} \\
\colhead{(1)} &
\colhead{(2)} &
\colhead{(3)} &
\colhead{(4)} &
\colhead{(5)} &
\colhead{(6)} &
\colhead{(7)} &
\colhead{(8)}  
}
\startdata
I & $41.8_{-4.1}^{+2.7}$ & $>0.992$ & 5682.7 & 5004 & 1.136 & 68 & 5819\\
II & $41.8_{-4.0}^{+3.2}$ & $>0.992$ & 5682.7 & 4996 & 1.137 & 72 & 5827\\
III & --- & --- & --- & --- & --- & --- & --- \\
IV & $33.2_{-6.5}^{+6.7}$ & $>0.996$ & 4677.5 & 4238 & 1.104 & --- & ---
\enddata
\tablecomments{
Col. (1): Spectral fit.
Col. (2): Inclination angle of the inner accretion disk.
Col. (3): Dimensionless BH spin.
Col. (4): Cash statistics.
Col. (5): Degree of freedom.
Col. (6): Reduced $C$-stat.
Col. (7): Number of parameters.
Col. (8): Akaike Information Criterion. In view of Fit~III yielding no results, a comparative analysis of the AIC between Fit~III and IV is omitted.
}
\label{tab:3c_soft}
\end{deluxetable}

\subsection{Effect of Soft Excess Interpretation}
\label{sec:soft_excess}

The interpretation of the soft excess has been a subject of ongoing debate \citep[e.g.,][]{crummy2006,done2012,jin2017,garcia2019,gliozzi2020}. The soft excess is, at least in part, a reflection feature that influences the determination of parameters such as $a_\ast$ \citep[e.g.,][]{walton2013}, ionization, and the emissivity profile. However, the soft band often introduces complications with complex absorption and scattering phenomena, mimicking the relativistic effects associated with reflection. Consequently, distinguishing whether the soft excess is due to reflection-based relativistic smearing or purely reprocessing poses a challenge. This ambiguity is often addressed using high-S/N broadband spectra \citep[e.g.,][]{guainazzi2006,risaliti2013,kammoun2018}. In \citet{ding2022} and our sections on observed spectra, the utility of multi-epoch, broadband, reflection-dominant spectra with well-modeled absorption reduces the complexities. 

Using 3C\,382 as an example, we then delve into a detailed discussion of the phenomenological modeling of the soft excess with little influence of the WA since WAs are well-modeled in Section~\ref{sec:modelfit}. As the \textsc{RelxillCp} model inadequately describes the soft excess in 3C\,382, we choose to model it using an additional phenomenological redshifted blackbody with a temperature tied among spectral epochs ($kT=8.7_{-0.3}^{+0.3} \times 10^{-2}\, \rm keV$). Despite statistical robustness, this introduces systematic uncertainty into the spin, ionization, and emissivity profile. The blackbody temperature may vary between epochs, and alternatively, adopting the \textsc{zBBody} model may introduce a selection bias. We conduct four spectral fits to elucidate the effect of our \textsc{zBBody} interpretation of the soft excess:

\begin{enumerate}[label=\Roman*.]
    \item The spectral fitting in Section~\ref{sec:obs_fitting}. 
    \item Untying blackbody temperatures between epochs.
    \item Ignoring spectral bins with energies lower than 2\,keV and directly producing the fit to check for the existence of a selection bias. 
    \item Ignoring spectral bins with energies lower than 2\,keV and deleting the absorber table components in the soft band. We then directly start an MCMC error calculation with the best-fit parameters from our original fit (Fit~I). 
\end{enumerate}

Results are presented in Table~\ref{tab:3c_soft}. The inclination $\theta_\mathrm{disk}$ and spin $a_\ast$ show no significant impact in Fit~II. Despite introducing four additional model parameters to account for the temperature, the $C$-stat remains unchanged ($\Delta C\text{-stat}=0.02$). The likelihood-ratio test cannot reject the null model (Fit~I), and this scenario is slightly disfavored with $\Delta \mathrm{AIC} = +8$. The blackbody temperatures across the five epochs align with the tied value in Fit~I ($|kT_\mathrm{a,b,c,d,e} - 0.087| <0.002$). Fit~III yields unattainable results when information in the soft band is completely lost. The values for inclination ($\theta_\mathrm{disk} \leq 3^\circ$), spin ($a_\ast = 0.998$), Fe abundance ($A_\mathrm{Fe}=10$), and most emissivity indices (7 out of 10 Indexes($_{1,2}$) $\to 10,0$) are all pegged to their boundaries. The reflection fraction ($R_f \lesssim 0.29$) is notably smaller than the values in Table~\ref{tab:3c_reflection_par}. Therefore, addressing the bias of using a specific soft excess model is challenging through a comparison of fits with broadband data and pure hard X-ray data. In Fit~IV, we abandon pursuing a fit and instead start error calculation on the known "best-fit" parameters directly. As expected, $a_\ast$, $A_\mathrm{Fe}$, and $\Gamma$ generally align with Fit~I, exhibiting overlapping 90\% error. However, $R_f$ is smaller ($<0.25$), and ionization is larger ($\log \xi \gtrsim 3.2$). The larger $\xi$ possibly shifts $\theta_\mathrm{disk}$ to a smaller value. Notably, \textsc{RELXILL} still fits the harder energies well in this scenario.

In summary, the epoch-independent blackbody effectively models the soft excess. We emphasize the importance of incorporating information from the soft band to enhance the precision of the reflection model fit.

\subsection{Data Rebinning}
\label{sec:stat}

Finally, we address a crucial technical aspect that can introduce biases in spectrum analysis---data rebinning. This process involves redividing and recombining data into a new set of bins. In spectroscopy, the decision to rebin over energy channels occurs after data reduction. While this reduces raw spectrum information, introduces potential errors, and increases the risk of creating artifacts, it is scientifically and statistically justified.

Handling broadband data from diverse instruments necessitates rebinning when assigning equal weights to all instruments. Analyzing spectra with significantly different photon counts, such as combining optical and X-ray spectra or XMM-Newton and NuSTAR spectra with varying X-ray emissions, can be challenging. High-count spectra would dominate the fit if the raw data were used without proper rebinning\footnote{Combining spectra from different instruments is generally discouraged due to differences in response matrices, which may introduce artifacts. Nevertheless, specific science applications may still benefit from this approach (e.g., \citealt{wilkins2021,ding2022}; see our data deduction for I\,Zwicky\,1 in Section~\ref{sec:data_reduc}).}. Moreover, even a single spectrum from one instrument may lose important information, like emission lines, if data bins are too narrow \citep[e.g.,][]{kaastra1999,kaastra2016}.

For statistical convenience, a common practice in X-ray observations is to rebin spectra so that each energy bin contains a minimum of 25 counts. This ensures approximately Gaussian data, facilitating interpretation with standard $\chi^2$ statistics. However, blindly adopting this approximation introduces bias in fitting \citep[e.g.,][]{humphrey2009,choudhury2017}. Poisson likelihood and Cash statistics are available for unbinned spectra in event-based X-ray observations. Our analysis of mock spectra (Section~4.2) uses Cash statistics. The debate on statistical choices originates from frequentist statistics, and introducing Bayesian approaches to X-ray spectrum analysis can alleviate it. The nested sampling method \citep{skilling2004}, widely embraced in cosmology \citep[e.g.,][]{planck2020} and gravitational wave studies \citep[e.g.,][]{abbott2019}, could be beneficial \citep[see, e.g.,][]{buchner2014}. This method offers advantages like model selection with the Bayes factor and robust parameter estimation. The nested sampling method is particularly useful for handling multiple modes/maxima in the posterior parameter distribution of reflection models, challenging for typical MCMC algorithms.

Rebinning introduces computational challenges, especially for instruments with high resolution or when the forward-folding process is repeated frequently in spectral fitting with many free parameters \citep[see detailed discussion in Section~2 of][]{kaastra2016}. This problem is magnified in upcoming high-resolution X-ray missions like XRISM \citep{tashiro2018} and Athena \citep{nandra2013}. While not rebinning the simulated spectra might extend MCMC analysis time, it helps avoid potential biases introduced during data reduction. Given that the simulations only involve a broad Fe line and lack complex emission lines, we believe oversampling does not compromise important information.

In summary, the choice of data binning involves a compromise between scientific and computational considerations. Balancing this tension often means rebining spectra to avoid oversampling the instrumental resolution by a factor of 3 (1.5 times the Nyquist rate; see \citealt{nyquist1928,shannon1949}). The application of systematic approaches like that proposed by \citet{kaastra2016}, considering both S/N and instrumental resolution, needs validation through tests on mock spectra to confirm its ability to reduce computational time while preserving critical spectral properties, especially complex reprocessing features. This remains a topic for future investigation.

\section{Conclusions}
\label{sec:conclusion}

We propose a systematic method to measure the inner accretion disk inclination of AGNs with broadband ($0.3-78\,\rm keV$) X-ray reflection spectroscopy. Using \textsc{RelxillCp} from the self-consistent reflection model family \textsc{RELXILL}, without assuming any particular coronal geometry, we test the reliability of measuring inclination by analyzing joint XMM-Newton and NuSTAR spectra of the narrow-line Seyfert~1 galaxy I\,Zwicky\,1 and the broad-line radio galaxy 3C\,382, alongside 13,860 simulated spectra that mimic joint XMM-Newton and NuSTAR observations of 3C\,382. 

Statistically consistent measurement of inclination can be achieved with \textsc{RelxillCp}. More accurate measurements can be obtained if multiple observations are combined and fitted simultaneously, and if the spectrum is more reflection-dominant. Multi-epoch, simultaneous data offer the best constraints on inner accretion disk inclination, but as long as prominent reflection features exist within the relevant instrumental coverage, inclination measurements can still be obtained, even for a single-epoch XMM-Newton observation. The presence of ionizing absorbers, albeit complicating the spectrum analysis process, in principle does not affect the inclination measurements. In the future, a systematic investigation of the impact of warm absorbers on X-ray spectral analysis would be worthwhile. At the current stage, case-by-case treatments of the soft band should be employed to help the fit converge and reduce systematic errors by successful modeling of the absorption.

Our mock tests indicate that systems with higher inclinations and larger reflection fractions recover the input inclinations more reliably. Systems with weaker reflection ($R_f \lesssim 0.3$) yield greater scatter in the inclination recovery, whereas for reflection-dominant systems ($R_f \gtrsim 5$) the precision improves mildly with a larger reflection fraction. The reflection fraction can be treated as a qualitative indicator of the systematic bias of applying \textsc{RelxillCp}. Higher iron abundance and coronal temperature help tighten the constraints on inner accretion disk inclination, but the improvements are not as significant as the reflection fraction. The spin of the BH, although degenerate with inclination, does not affect significantly the measurement of inclination.

\section*{Acknowledgements}
LCH was supported by the National Science Foundation of China (11721303, 11991052, 12011540375, 12233001), the National Key R\&D Program of China (2022YFF0503401), and the China Manned Space Project (CMS-CSST-2021-A04, CMS-CSST-2021-A06).  We thank the anonymous referee for helpful suggestions. RD is grateful to Cosimo Bambi, James F. Steiner, and Javier A. Garc\'ia for useful suggestions in the early stages of the paper, Kishalay Choudhury for explaining the details of \citet{choudhury2017}, Elias S. Kammoun for explaining \citet{kammoun2018} and sharing their script on resolution binning, and Shuo Feng for advice on writing. 

All figures in this paper are produced with \textsc{SciencePlots} \citep{SciencePlots}. Other computational platforms and packages used include: \textsc{AstroPy} \citep{astropy2013,astropy2018,astropy2022}, \textsc{Corner} \citep{foremanmackey2016}, \textsc{Numpy} \citep{walt2011,harris2020}, and \textsc{Matplotlib} \citep{hunter2007}.

\bibliographystyle{aasjournal}


\begin{thebibliography}{}
\expandafter\ifx\csname natexlab\endcsname\relax\def\natexlab#1{#1}\fi

\bibitem[{{Abbott} {et~al.}(2019)}]{abbott2019}
{Abbott}, B.~P., {Abbott}, R., {Abbott}, T.~D., {et~al.} 2019, \apjl, 882, L24

\bibitem[Akaike(1974)]{akaike1974}
Akaike, H.\ 1974, IEEE Transactions on Automatic Control, 19, 716

\bibitem[{{Antonucci}(1993)}]{antonucci1993}
{Antonucci}, R. 1993, \araa, 31, 473

\bibitem[{{Arnaud}(1996)}]{arnaud1996}
{Arnaud}, K.~A. 1996, in ASP Conf. Ser., Vol. 101, Astronomical Data Analysis Software and Systems V, ed. G.~H. {Jacoby} \& J.~{Barnes} (San Francisco, CA: ASP), 17

\bibitem[{{Arnaud}(2016)}]{arnaud2016}
{Arnaud}, K.~A. 2016, HEAD, 115.02

\bibitem[{{Astropy Collaboration} {et~al.}(2022)}]{astropy2022}
{Astropy Collaboration}, {Price-Whelan}, A.~M., {Lim}, P.~L., {et~al.} 2022, \apj, 935, 167

\bibitem[{{Astropy Collaboration} {et~al.}(2018)}]{astropy2018}
{Astropy Collaboration}, {Price-Whelan}, A.~M., {Sip{\H{o}}cz}, B.~M., {et~al.} 2018, \aj, 156, 123

\bibitem[{{Astropy Collaboration} {et~al.}(2013){Astropy Collaboration},
  {Robitaille}, {Tollerud}, {Greenfield}, {Droettboom}, {Bray}, {Aldcroft},
  {Davis}, {Ginsburg}, {Price-Whelan}, {Kerzendorf}, {Conley}, {Crighton},
  {Barbary}, {Muna}, {Ferguson}, {Grollier}, {Parikh}, {Nair}, {Unther},
  {Deil}, {Woillez}, {Conseil}, {Kramer}, {Turner}, {Singer}, {Fox}, {Weaver},
  {Zabalza}, {Edwards}, {Azalee Bostroem}, {Burke}, {Casey}, {Crawford},
  {Dencheva}, {Ely}, {Jenness}, {Labrie}, {Lim}, {Pierfederici}, {Pontzen},
  {Ptak}, {Refsdal}, {Servillat}, \& {Streicher}}]{astropy2013}
{Astropy Collaboration}, {Robitaille}, T.~P., {Tollerud}, E.~J., {et~al.} 2013, \aap, 558, A33

\bibitem[{{Ballantyne} {et~al.}(2014){Ballantyne}, {Bollenbacher}, {Brenneman},
  {Madsen}, {Balokovi{\'c}}, {Boggs}, {Christensen}, {Craig}, {Gandhi},
  {Hailey}, {Harrison}, {Lohfink}, {Marinucci}, {Markwardt}, {Stern}, {Walton},
  \& {Zhang}}]{ballantyne2014}
{Ballantyne}, D.~R., {Bollenbacher}, J.~M., {Brenneman}, L.~W., {et~al.} 2014,
  \apj, 794, 62

\bibitem[Balokovi{\'c} et al.(2020)]{balokovic2020} 
Balokovi{\'c}, M., Harrison, F.~A., Madejski, G., et al.\ 2020, \apj, 905, 41. 

\bibitem[{{Bardeen} \& {Petterson}(1975)}]{bardeen1975}
{Bardeen}, J.~M., \& {Petterson}, J.~A. 1975, \apjl, 195, L65

\bibitem[{{Baskin} \& {Laor}(2018)}]{baskin2018}
{Baskin}, A., \& {Laor}, A. 2018, \mnras, 474, 1970

\bibitem[{{Bautista} \& {Kallman}(2001)}]{bautista2001}
{Bautista}, M.~A., \& {Kallman}, T.~R. 2001, \apjs, 134, 139

\bibitem[{{Belmont} {et~al.}(2008){Belmont}, {Malzac}, \&
  {Marcowith}}]{belmont2008}
{Belmont}, R., {Malzac}, J., \& {Marcowith}, A. 2008, \aap, 491, 617

\bibitem[{{Beuchert} {et~al.}(2017){Beuchert}, {Markowitz}, {Dauser},
  {Garc{\'\i}a}, {Keck}, {Wilms}, {Kadler}, {Brenneman}, \&
  {Zdziarski}}]{beuchert2017}
{Beuchert}, T., {Markowitz}, A.~G., {Dauser}, T., {et~al.} 2017, \aap, 603, A50

\bibitem[Birkinshaw \& Davies(1985)]{birkinshaw1985}
Birkinshaw, M. \& Davies, R.~L.\ 1985, \apj, 291, 32

\bibitem[{{Blandford} \& {McKee}(1982)}]{blandford1982}
{Blandford}, R.~D., \& {McKee}, C.~F. 1982, \apj, 255, 419

\bibitem[{{Blandford} \& {Znajek}(1977)}]{blandford1977}
{Blandford}, R.~D., \& {Znajek}, R.~L. 1977, \mnras, 179, 433

\bibitem[{{Bonson} \& {Gallo}(2016)}]{bonson2016}
{Bonson}, K., \& {Gallo}, L.~C. 2016, \mnras, 458, 1927

\bibitem[{{Brenneman} \& {Reynolds}(2006)}]{brenneman2006}
{Brenneman}, L.~W., \& {Reynolds}, C.~S. 2006, \apj, 652, 1028

\bibitem[{{Buchner} {et~al.}(2014){Buchner}, {Georgakakis}, {Nandra}, {Hsu},
  {Rangel}, {Brightman}, {Merloni}, {Salvato}, {Donley}, \&
  {Kocevski}}]{buchner2014}
{Buchner}, J., {Georgakakis}, A., {Nandra}, K., {et~al.} 2014, \aap, 564, A125

\bibitem[{{Caballero-Garc{\'\i}a} {et~al.}(2020){Caballero-Garc{\'\i}a},
  {Papadakis}, {Dov{\v{c}}iak}, {Bursa}, {Svoboda}, \&
  {Karas}}]{caballerogarcia2020}
{Caballero-Garc{\'\i}a}, M.~D., {Papadakis}, I.~E., {Dov{\v{c}}iak}, M.,
  {et~al.} 2020, \mnras, 498, 3184

\bibitem[{{Cash}(1979)}]{cash1979}
{Cash}, W. 1979, \apj, 228, 939

\bibitem[{{Choudhury} {et~al.}(2017){Choudhury}, {Garc{\'\i}a}, {Steiner}, \&
  {Bambi}}]{choudhury2017}
{Choudhury}, K., {Garc{\'\i}a}, J.~A., {Steiner}, J.~F., \& {Bambi}, C. 2017,
  \apj, 851, 57

\bibitem[{{Compton}(1923)}]{compton1923}
{Compton}, A.~H. 1923, Physical Review, 21, 483

\bibitem[Costantini et~al.(2007)]{costantini2007}
{Costantini}, E., {Gallo}, L.~C., {Brandt}, W.~N., {Fabian}, A.~C., \&
  {Boller}, T. 2007, \mnras, 378, 873

\bibitem[Crummy et al.(2006)]{crummy2006}
Crummy, J., Fabian, A.~C., Gallo, L., et al.\ 2006, \mnras, 365, 1067

\bibitem[{{Czerny} {et~al.}(2016){Czerny}, {Du}, {Wang}, \&
  {Karas}}]{czerny2016}
{Czerny}, B., {Du}, P., {Wang}, J.-M., \& {Karas}, V. 2016, \apj, 832, 15

\bibitem[{{Czerny} \& {Hryniewicz}(2011)}]{czerny2011blr}
{Czerny}, B., \& {Hryniewicz}, K. 2011, \aap, 525, L8

\bibitem[Czerny et~al.(2011)]{czerny2011} Czerny, B., Hryniewicz, K., Niko\l ajuk, M., \& Sadowski, A. 2011, \mnras, 415, 2942

\bibitem[Czerny et al.(2017)]{czerny2017}
Czerny, B., Li, Y.-R., Hryniewicz, K., et al.\ 2017, \apj, 846, 154

\bibitem[Dauser et al.(2022)]{dauser2022}
Dauser, T., Garc{\'\i}a, J.~A., Joyce, A., et al.\ 2022, \mnras, 514, 3965

\bibitem[{{Dauser} {et~al.}(2014){Dauser}, {Garcia}, {Parker}, {Fabian}, \&
  {Wilms}}]{dauser2014}
{Dauser}, T., {Garcia}, J., {Parker}, M.~L., {Fabian}, A.~C., \& {Wilms}, J.
  2014, \mnras, 444, L100

\bibitem[{{Dauser} {et~al.}(2016){Dauser}, {Garc{\'\i}a}, {Walton}, {Eikmann},
  {Kallman}, {McClintock}, \& {Wilms}}]{dauser2016}
{Dauser}, T., {Garc{\'\i}a}, J., {Walton}, D.~J., {et~al.} 2016, \aap, 590, A76

\bibitem[{{Dauser} {et~al.}(2013){Dauser}, {Garcia}, {Wilms}, {B{\"o}ck},
  {Brenneman}, {Falanga}, {Fukumura}, \& {Reynolds}}]{dauser2013}
{Dauser}, T., {Garcia}, J., {Wilms}, J., {et~al.} 2013, \mnras, 430, 1694

\bibitem[{{Dauser} {et~al.}(2012){Dauser}, {Svoboda}, {Schartel}, {Wilms},
  {Dov{\v{c}}iak}, {Ehle}, {Karas}, {Santos-Lle{\'o}}, \&
  {Marshall}}]{dauser2012}
{Dauser}, T., {Svoboda}, J., {Schartel}, N., {et~al.} 2012, \mnras, 422, 1914

\bibitem[{{Dauser} {et~al.}(2010){Dauser}, {Wilms}, {Reynolds}, \&
  {Brenneman}}]{dauser2010}
{Dauser}, T., {Wilms}, J., {Reynolds}, C.~S., \& {Brenneman}, L.~W. 2010,
  \mnras, 409, 1534

\bibitem[{{den Herder} {et~al.}(2001)}]{den-herder2001}
{den Herder}, J.~W., {Brinkman}, A.~C., {Kahn}, S.~M., {et~al.} 2001, \aap, 365, L7

\bibitem[{{Ding} {et~al.}(2022){Ding}, {Li}, {Ho}, \& {Ricci}}]{ding2022}
{Ding}, Y., {Li}, R., {Ho}, L.~C., \& {Ricci}, C. 2022, \apj, 931, 77

\bibitem[Done et al.(2012)]{done2012}
Done, C., Davis, S.~W., Jin, C., et al.\ 2012, \mnras, 420, 1848

\bibitem[Done et~al.(2013)]{done2013}
{Done}, C., {Jin}, C., {Middleton}, M., \& {Ward}, M. 2013, \mnras, 434, 1955

\bibitem[Done et al.(2007)]{done2007}
Done, C., Sobolewska, M.~A., Gierli{\'n}ski, M., et al.\ 2007, \mnras, 374, L15

\bibitem[Dov{\v{c}}iak et al.(2004)]{dovciak2004}
Dov{\v{c}}iak, M., Karas, V., \& Yaqoob, T.\ 2004, \apjs, 153, 205

\bibitem[Dov{\v{c}}iak et al.(2022)]{dovciak2022}
Dov{\v{c}}iak, M., Papadakis, I.~E., Kammoun, E.~S., et al.\ 2022, \aap, 661, A135

\bibitem[{{Dove} {et~al.}(1997){Dove}, {Wilms}, {Maisack}, \&
  {Begelman}}]{dove1997}
{Dove}, J.~B., {Wilms}, J., {Maisack}, M., \& {Begelman}, M.~C. 1997, \apj,
  487, 759

\bibitem[{{Einstein}(1916)}]{einstein1916}
{Einstein}, A. 1916, Annalen der Physik, 354, 769

\bibitem[Emmering et al.(1992)]{emmering1992}
Emmering, R.~T., Blandford, R.~D., \& Shlosman, I.\ 1992, \apj, 385, 460

\bibitem[{{Eracleous} \& {Halpern}(1994)}]{eracleous1994}
{Eracleous}, M., \& {Halpern}, J.~P. 1994, \apjs, 90, 1

\bibitem[Fabian(2006)]{fabian2006}
Fabian, A.~C.\ 2006, Astronomische Nachrichten, 327, 943

\bibitem[{{Fabian} {et~al.}(2015){Fabian}, {Lohfink}, {Kara}, {Parker},
  {Vasudevan}, \& {Reynolds}}]{fabian2015}
{Fabian}, A.~C., {Lohfink}, A., {Kara}, E., {et~al.} 2015, \mnras, 451, 4375

\bibitem[{{Fabian} {et~al.}(1989){Fabian}, {Rees}, {Stella}, \&
  {White}}]{fabian1989}
{Fabian}, A.~C., {Rees}, M.~J., {Stella}, L., \& {White}, N.~E. 1989, \mnras,
  238, 729

\bibitem[{{Fabian} {et~al.}(2002){Fabian}, {Vaughan}, {Nandra}, {Iwasawa},
  {Ballantyne}, {Lee}, {De Rosa}, {Turner}, \& {Young}}]{fabian2002}
{Fabian}, A.~C., {Vaughan}, S., {Nandra}, K., {et~al.} 2002, \mnras, 335, L1

\bibitem[{{Fabian} {et~al.}(2009){Fabian}, {Zoghbi}, {Ross}, {Uttley}, {Gallo},
  {Brandt}, {Blustin}, {Boller}, {Caballero-Garcia}, {Larsson}, {Miller},
  {Miniutti}, {Ponti}, {Reis}, {Reynolds}, {Tanaka}, \& {Young}}]{fabian2009}
{Fabian}, A.~C., {Zoghbi}, A., {Ross}, R.~R., {et~al.} 2009, \nat, 459, 540

\bibitem[{{Fischer} {et~al.}(2013){Fischer}, {Crenshaw}, {Kraemer}, \&
  {Schmitt}}]{fischer2013}
{Fischer}, T.~C., {Crenshaw}, D.~M., {Kraemer}, S.~B., \& {Schmitt}, H.~R.
  2013, \apjs, 209, 1

\bibitem[{{Fischer} {et~al.}(2014){Fischer}, {Crenshaw}, {Kraemer}, {Schmitt},
  \& {Turner}}]{fischer2014}
{Fischer}, T.~C., {Crenshaw}, D.~M., {Kraemer}, S.~B., {Schmitt}, H.~R., \&
  {Turner}, T.~J. 2014, \apj, 785, 25

\bibitem[{{Foreman-Mackey}(2016)}]{foremanmackey2016}
{Foreman-Mackey}, D. 2016, The Journal of Open Source Software, 1, 24

\bibitem[{{Foreman-Mackey} {et~al.}(2013){Foreman-Mackey}, {Hogg}, {Lang}, \&
  {Goodman}}]{foremanmackey2013}
{Foreman-Mackey}, D., {Hogg}, D.~W., {Lang}, D., \& {Goodman}, J. 2013, \pasp,
  125, 306

\bibitem[F{\"u}rst et al.(2016)]{furst2016}
F{\"u}rst, F., M{\"u}ller, C., Madsen, K.~K., et al.\ 2016, \apj, 819, 150

\bibitem[Gallo et al.(2004)]{gallo2004}
Gallo, L.~C., Boller, T., Brandt, W.~N., et al.\ 2004, \aap, 417, 29

\bibitem[{{Garc{\'\i}a} {et~al.}(2014){Garc{\'\i}a}, {Dauser}, {Lohfink},
  {Kallman}, {Steiner}, {McClintock}, {Brenneman}, {Wilms}, {Eikmann},
  {Reynolds}, \& {Tombesi}}]{garcia2014}
{Garc{\'\i}a}, J., {Dauser}, T., {Lohfink}, A., {et~al.} 2014, \apj, 782, 76

\bibitem[{{Garc{\'\i}a} {et~al.}(2013){Garc{\'\i}a}, {Dauser}, {Reynolds},
  {Kallman}, {McClintock}, {Wilms}, \& {Eikmann}}]{garcia2013}
{Garc{\'\i}a}, J., {Dauser}, T., {Reynolds}, C.~S., {et~al.} 2013, \apj, 768,
  146

\bibitem[{{Garc{\'\i}a} \& {Kallman}(2010)}]{garcia2010}
{Garc{\'\i}a}, J., \& {Kallman}, T.~R. 2010, \apj, 718, 695

\bibitem[{{Garc{\'\i}a} {et~al.}(2011){Garc{\'\i}a}, {Kallman}, \&
  {Mushotzky}}]{garcia2011}
{Garc{\'\i}a}, J., {Kallman}, T.~R., \& {Mushotzky}, R.~F. 2011, \apj, 731, 131

\bibitem[Garc{\'\i}a et al.(2019)]{garcia2019}
Garc{\'\i}a, J.~A., Kara, E., Walton, D., et al.\ 2019, \apj, 871, 88

\bibitem[{Garrett(2021)}]{SciencePlots}
Garrett, J. 2021, SciencePlots (v1.0.9), v.1.0.9,  Zenodo, doi:10.5281/zenodo.5512926

\bibitem[{{Ghisellini} {et~al.}(1993){Ghisellini}, {Padovani}, {Celotti}, \&
  {Maraschi}}]{ghisellini1993}
{Ghisellini}, G., {Padovani}, P., {Celotti}, A., \& {Maraschi}, L. 1993, \apj,
  407, 65
  
\bibitem[Gliozzi \& Williams(2020)]{gliozzi2020}
Gliozzi, M. \& Williams, J.~K.\ 2020, \mnras, 491, 532

\bibitem[{{Goodman} \& {Weare}(2010)}]{goodman2010}
{Goodman}, J., \& {Weare}, J. 2010, Communications in Applied Mathematics and
  Computational Science, 5, 65

\bibitem[Greenhill et al.(2009)]{greenhill2009}
Greenhill, L.~J., Kondratko, P.~T., Moran, J.~M., et al.\ 2009, \apj, 707, 787

\bibitem[Guainazzi et al.(2006)]{guainazzi2006}
Guainazzi, M., Bianchi, S., \& Dov{\v{c}}iak, M.\ 2006, Astronomische Nachrichten, 327, 1032

\bibitem[{{Guilbert} \& {Rees}(1988)}]{guilbert1988}
{Guilbert}, P.~W., \& {Rees}, M.~J. 1988, \mnras, 233, 475

\bibitem[{{Haardt} \& {Maraschi}(1991)}]{haardt1991}
{Haardt}, F., \& {Maraschi}, L. 1991, \apjl, 380, L51

\bibitem[{{Haardt} \& {Maraschi}(1993)}]{haardt1993}
{Haardt}, F., \& {Maraschi}, L. 1993, \apj, 413, 507

\bibitem[{{Harris} {et~al.}(2020){Harris}, {Millman}, {van der Walt},
  {Gommers}, {Virtanen}, {Cournapeau}, {Wieser}, {Taylor}, {Berg}, {Smith},
  {Kern}, {Picus}, {Hoyer}, {van Kerkwijk}, {Brett}, {Haldane}, {del R{\'\i}o},
  {Wiebe}, {Peterson}, {G{\'e}rard-Marchant}, {Sheppard}, {Reddy}, {Weckesser},
  {Abbasi}, {Gohlke}, \& {Oliphant}}]{harris2020}
{Harris}, C.~R., {Millman}, K.~J., {van der Walt}, S.~J., {et~al.} 2020, \nat,
  585, 357

\bibitem[{{Harrison} {et~al.}(2013){Harrison}, {Craig}, {Christensen},
  {Hailey}, {Zhang}, {Boggs}, {Stern}, {Cook}, {Forster}, {Giommi},
  {Grefenstette}, {Kim}, {Kitaguchi}, {Koglin}, {Madsen}, {Mao}, {Miyasaka},
  {Mori}, {Perri}, {Pivovaroff}, {Puccetti}, {Rana}, {Westergaard}, {Willis},
  {Zoglauer}, {An}, {Bachetti}, {Barri{\`e}re}, {Bellm}, {Bhalerao},
  {Brejnholt}, {Fuerst}, {Liebe}, {Markwardt}, {Nynka}, {Vogel}, {Walton},
  {Wik}, {Alexander}, {Cominsky}, {Hornschemeier}, {Hornstrup}, {Kaspi},
  {Madejski}, {Matt}, {Molendi}, {Smith}, {Tomsick}, {Ajello}, {Ballantyne},
  {Balokovi{\'c}}, {Barret}, {Bauer}, {Blandford}, {Brandt}, {Brenneman},
  {Chiang}, {Chakrabarty}, {Chenevez}, {Comastri}, {Dufour}, {Elvis}, {Fabian},
  {Farrah}, {Fryer}, {Gotthelf}, {Grindlay}, {Helfand}, {Krivonos}, {Meier},
  {Miller}, {Natalucci}, {Ogle}, {Ofek}, {Ptak}, {Reynolds}, {Rigby},
  {Tagliaferri}, {Thorsett}, {Treister}, \& {Urry}}]{harrison2013}
{Harrison}, F.~A., {Craig}, W.~W., {Christensen}, F.~E., {et~al.} 2013, \apj,
  770, 103

\bibitem[{{Herrnstein} {et~al.}(1999){Herrnstein}, {Moran}, {Greenhill},
  {Diamond}, {Inoue}, {Nakai}, {Miyoshi}, {Henkel}, \&
  {Riess}}]{herrnstein1999}
{Herrnstein}, J.~R., {Moran}, J.~M., {Greenhill}, L.~J., {et~al.} 1999, \nat,
  400, 539

\bibitem[{{Hicks} \& {Malkan}(2008)}]{hicks2008}
{Hicks}, E. K.~S., \& {Malkan}, M.~A. 2008, \apjs, 174, 31

\bibitem[{{Ho} {et~al.}(2000){Ho}, {Rudnick}, {Rix}, {Shields}, {McIntosh},
  {Filippenko}, {Sargent}, \& {Eracleous}}]{ho2000}
{Ho}, L.~C., {Rudnick}, G., {Rix}, H.-W., {et~al.} 2000, \apj, 541, 120

\bibitem[{{H{\"o}nig} {et~al.}(2007)}]{honig2007}
{H{\"o}nig}, S.~F., {Beckert}, T., {Ohnaka}, K., \& {Weigelt}, G. 2007, in The Central Engine of Active Galactic Nuclei, ed. L.~C. {Ho} \& J.-M. {Wang} (San Francisco, CA: ASP), 487

\bibitem[{{Hopkins} {et~al.}(2012){Hopkins}, {Hernquist}, {Hayward}, \&
  {Narayanan}}]{hopkins2012}
{Hopkins}, P.~F., {Hernquist}, L., {Hayward}, C.~C., \& {Narayanan}, D. 2012,
  \mnras, 425, 1121

\bibitem[{{Humphrey} {et~al.}(2009){Humphrey}, {Liu}, \&
  {Buote}}]{humphrey2009}
{Humphrey}, P.~J., {Liu}, W., \& {Buote}, D.~A. 2009, \apj, 693, 822

\bibitem[{{Hunter}(2007)}]{hunter2007}
{Hunter}, J.~D. 2007, Computing in Science and Engineering, 9, 90

\bibitem[Ingram et al.(2023)]{ingram2023}
Ingram, A., Ewing, M., Marinucci, A., et al.\ 2023, \mnras, 525, 5437

\bibitem[Ingram et al.(2019)]{ingram2019}
Ingram, A., Mastroserio, G., Dauser, T., et al.\ 2019, \mnras, 488, 324

\bibitem[{{Jansen} {et~al.}(2001){Jansen}, {Lumb}, {Altieri}, {Clavel}, {Ehle},
  {Erd}, {Gabriel}, {Guainazzi}, {Gondoin}, {Much}, {Munoz}, {Santos},
  {Schartel}, {Texier}, \& {Vacanti}}]{jansen2001}
{Jansen}, F., {Lumb}, D., {Altieri}, B., {et~al.} 2001, \aap, 365, L1

\bibitem[{{Jiang} {et~al.}(2019){Jiang}, {Blaes}, {Stone}, \&
  {Davis}}]{jiang2019}
{Jiang}, Y.-F., {Blaes}, O., {Stone}, J.~M., \& {Davis}, S.~W. 2019, \apj, 885, 144

\bibitem[Jin et al.(2017)]{jin2017}
Jin, C., Done, C., \& Ward, M.\ 2017, \mnras, 468, 3663

\bibitem[{{Jin} {et~al.}(2012){Jin}, {Ward}, {Done}, \& {Gelbord}}]{jin2012}
{Jin}, C., {Ward}, M., {Done}, C., \& {Gelbord}, J. 2012, \mnras, 420, 1825

\bibitem[Kaastra et al.(1996)]{kaastra1996}
Kaastra, J.~S., Mewe, R., \& Nieuwenhuijzen, H.\ 1996, UV and X-ray Spectroscopy of Astrophysical and Laboratory Plasmas, 411

\bibitem[{{Kaastra}(1999)}]{kaastra1999}
{Kaastra}, J.~S. 1999, in X-ray Spectroscopy in Astrophysics, Vol. 520, ed. J.~{van Paradijs} \& J.~A.~M. {Bleeker} (Berlin and Heidelberg, Germany: Springer), 269

\bibitem[{{Kaastra} \& {Bleeker}(2016)}]{kaastra2016}
{Kaastra}, J.~S., \& {Bleeker}, J.~A.~M. 2016, \aap, 587, A151

\bibitem[{{Kalberla} {et~al.}(2005){Kalberla}, {Burton}, {Hartmann}, {Arnal},
  {Bajaja}, {Morras}, \& {P{\"o}ppel}}]{kalberla2005}
{Kalberla}, P.~M.~W., {Burton}, W.~B., {Hartmann}, D., {et~al.} 2005, \aap,
  440, 775

\bibitem[{{Kallman} \& {Bautista}(2001)}]{kallman2001}
{Kallman}, T., \& {Bautista}, M. 2001, \apjs, 133, 221

\bibitem[{{Kammoun} {et~al.}(2018){Kammoun}, {Nardini}, \&
  {Risaliti}}]{kammoun2018}
{Kammoun}, E.~S., {Nardini}, E., \& {Risaliti}, G. 2018, \aap, 614, A44

\bibitem[Kamraj et al.(2022)]{kamraj2022} Kamraj, N., Brightman, M., Harrison, F.~A., et al.\ 2022, \apj, 927, 42. 

\bibitem[{{Kara} {et~al.}(2016){Kara}, {Alston}, {Fabian}, {Cackett}, {Uttley},
  {Reynolds}, \& {Zoghbi}}]{kara2016}
{Kara}, E., {Alston}, W.~N., {Fabian}, A.~C., {et~al.} 2016, \mnras, 462, 511

\bibitem[{{Kerr}(1963)}]{kerr1963}
{Kerr}, R.~P. 1963, \prl, 11, 237

\bibitem[{{Kinch} {et~al.}(2020){Kinch}, {Noble}, {Schnittman}, \&
  {Krolik}}]{kinch2020}
{Kinch}, B.~E., {Noble}, S.~C., {Schnittman}, J.~D., \& {Krolik}, J.~H. 2020,
  \apj, 904, 117

\bibitem[{{Kinch} {et~al.}(2016){Kinch}, {Schnittman}, {Kallman}, \&
  {Krolik}}]{kinch2016}
{Kinch}, B.~E., {Schnittman}, J.~D., {Kallman}, T.~R., \& {Krolik}, J.~H. 2016, \apj, 826, 52

\bibitem[King et al.(2005)]{king2005}
King, A.~R., Lubow, S.~H., Ogilvie, G.~I., et al.\ 2005, \mnras, 363, 49


\bibitem[Kinney et al.(2000)]{kinney2000}
Kinney, A.~L., Schmitt, H.~R., Clarke, C.~J., et al.\ 2000, \apj, 537, 152

\bibitem[Klein \& Nishina(1929)]{klein1929}
{Klein}, O., \& {Nishina}, T. 1929, Zeitschrift fur Physik, 52, 853

\bibitem[Komissarov(2022)]{komissarov2022}
Komissarov, S.~S.\ 2022, \mnras, 512, 2798

\bibitem[Kormendy \& Ho(2013)]{kormendy2013}
Kormendy, J., \& Ho, L.~C.\ 2013, \araa, 51, 511

\bibitem[Krolik(1998)]{krolik1998} Krolik, J. H. 1998, Active Galactic Nuclei (Princeton: Princeton Univ. Press)

\bibitem[{{Laor}(1991)}]{laor1991}
{Laor}, A. 1991, \apj, 376, 90

\bibitem[Li et al.(2018)]{li2018}
Li, Y.-R., Songsheng, Y.-Y., Qiu, J., et al.\ 2018, \apj, 869, 137

\bibitem[Li et al.(2015)]{li2015}
Li, Y.-R., Wang, J.-M., Cheng, C., et al.\ 2015, \apj, 804, 45

\bibitem[{{Lightman} \& {White}(1988)}]{lightman1988}
{Lightman}, A.~P., \& {White}, T.~R. 1988, \apj, 335, 57

\bibitem[Liska et al.(2021)]{liska2021}
Liska, M., Hesp, C., Tchekhovskoy, A., et al.\ 2021, \mnras, 507, 983

\bibitem[{{Magdziarz} \& {Zdziarski}(1995)}]{magdziarz1995}
{Magdziarz}, P., \& {Zdziarski}, A.~A. 1995, \mnras, 273, 837

\bibitem[{{Martocchia} \& {Matt}(1996)}]{martocchia1996}
{Martocchia}, A., \& {Matt}, G. 1996, \mnras, 282, L53

\bibitem[Mastroserio et al.(2021)]{mastroserio2021}
Mastroserio, G., Ingram, A., Wang, J., et al.\ 2021, \mnras, 507, 55

\bibitem[{{Matt} {et~al.}(1991){Matt}, {Perola}, \& {Piro}}]{matt1991}
{Matt}, G., {Perola}, G.~C., \& {Piro}, L. 1991, \aap, 247, 25

\bibitem[{{Metropolis} {et~al.}(1953){Metropolis}, {Rosenbluth}, {Rosenbluth},
  {Teller}, \& {Teller}}]{metropolis1953}
{Metropolis}, N., {Rosenbluth}, A.~W., {Rosenbluth}, M.~N., {Teller}, A.~H., \& {Teller}, E. 1953, \jcp, 21, 1087

\bibitem[{{Middleton} {et~al.}(2016){Middleton}, {Parker}, {Reynolds},
  {Fabian}, \& {Lohfink}}]{middleton2016}
{Middleton}, M.~J., {Parker}, M.~L., {Reynolds}, C.~S., {Fabian}, A.~C., \&
  {Lohfink}, A.~M. 2016, \mnras, 457, 1568

\bibitem[{{Miller} {et~al.}(2015){Miller}, {Tomsick}, {Bachetti}, {Wilkins},
  {Boggs}, {Christensen}, {Craig}, {Fabian}, {Grefenstette}, {Hailey},
  {Harrison}, {Kara}, {King}, {Stern}, \& {Zhang}}]{miller2015}
{Miller}, J.~M., {Tomsick}, J.~A., {Bachetti}, M., {et~al.} 2015, \apjl, 799, L6

\bibitem[{{Miniutti} \& {Fabian}(2004)}]{miniutti2004}
{Miniutti}, G., \& {Fabian}, A.~C. 2004, \mnras, 349, 1435

\bibitem[Misner et al.(1973)]{misner1973} 
Misner, C.~W., Thorne, K.~S., \& Wheeler, J.~A.\ 1973, Gravitation (San Francisco, CA: W. H. Freeman and Company)

\bibitem[Murray et al.(1995)]{murray1995}
Murray, N., Chiang, J., Grossman, S.~A., et al.\ 1995, \apj, 451, 498

\bibitem[{{Nandra} {et~al.}(2013)}]{nandra2013} 
{Nandra}, K., {Barret}, D., {Barcons}, X., {et~al.} 2013, arXiv:1306.2307

\bibitem[{{Natarajan} \& {Pringle}(1998)}]{natarajan1998}
{Natarajan}, P., \& {Pringle}, J.~E. 1998, \apjl, 506, L97

\bibitem[{{Newman} {et~al.}(1965){Newman}, {Couch}, {Chinnapared}, {Exton},
  {Prakash}, \& {Torrence}}]{newman1965}
{Newman}, E.~T., {Couch}, E., {Chinnapared}, K., {et~al.} 1965, Journal of
  Mathematical Physics, 6, 918

\bibitem[Nied{\'z}wiecki et al.(2019)]{niedzwiecki2019}
Nied{\'z}wiecki, A., Szanecki, M., \& Zdziarski, A.~A.\ 2019, \mnras, 485, 2942

\bibitem[{{Novikov} \& {Thorne}(1973)}]{novikov1973}
{Novikov}, I.~D., \& {Thorne}, K.~S. 1973, in Black Holes (Les Astres Occlus), ed. C. DeWitt \& B. DeWitt (New York: Gordon and Breach), 343

\bibitem[{{Nyquist}(1928)}]{nyquist1928}
{Nyquist}, H. 1928, Physical Review, 32, 110

\bibitem[{{Pancoast} {et~al.}(2014){Pancoast}, {Brewer}, {Treu}, {Park},
  {Barth}, {Bentz}, \& {Woo}}]{pancoast2014}
{Pancoast}, A., {Brewer}, B.~J., {Treu}, T., {et~al.} 2014, \mnras, 445, 3073

\bibitem[{{Papaloizou} \& {Lin}(1995)}]{papaloizou1995}
{Papaloizou}, J.~C.~B., \& {Lin}, D.~N.~C. 1995, \apj, 438, 841

\bibitem[{{Petrucci} {et~al.}(2018){Petrucci}, {Ursini}, {De Rosa}, {Bianchi},
  {Cappi}, {Matt}, {Dadina}, \& {Malzac}}]{petrucci2018}
{Petrucci}, P.~O., {Ursini}, F., {De Rosa}, A., {et~al.} 2018, \aap, 611, A59

\bibitem[Pjanka et al.(2017)]{pjanka2017}
Pjanka, P., Greene, J.~E., Seth, A.~C., et al.\ 2017, \apj, 844, 165

\bibitem[{{Planck Collaboration} {et~al.}(2020)}]{planck2020}
{Planck Collaboration}, {Akrami}, Y., {Arroja}, F., {et~al.} 2020, \aap, 641, A10

\bibitem[Ponti et al.(2018)]{ponti2018}
Ponti, G., Bianchi, S., Mu{\~n}oz-Darias, T., et al.\ 2018, \mnras, 473, 2304

\bibitem[{{Pringle}(1992)}]{pringle1992}
{Pringle}, J.~E. 1992, \mnras, 258, 811

\bibitem[{{Rees}(1966)}]{rees1966}
{Rees}, M.~J. 1966, \nat, 211, 468

\bibitem[{{Reeves} \& {Braito}(2019)}]{reeves2019}
{Reeves}, J.~N., \& {Braito}, V. 2019, \apj, 884, 80

\bibitem[{{Reynolds}(2014)}]{reynolds2014}
{Reynolds}, C.~S. 2014, \ssr, 183, 277

\bibitem[{{Reynolds}(2021)}]{reynolds2021}
{Reynolds}, C.~S. 2021, \araa, 59, 117

\bibitem[{{Reynolds} \& {Begelman}(1997)}]{reynolds1997}
{Reynolds}, C.~S., \& {Begelman}, M.~C. 1997, \apj, 488, 109

\bibitem[{{Reynolds} {et~al.}(1999){Reynolds}, {Young}, {Begelman}, \&
  {Fabian}}]{reynolds1999}
{Reynolds}, C.~S., {Young}, A.~J., {Begelman}, M.~C., \& {Fabian}, A.~C. 1999,
  \apj, 514, 164

\bibitem[Ricci et al.(2017)]{ricci2017} Ricci, C., Trakhtenbrot, B., Koss, M.~J., et al.\ 2017, \apjs, 233, 17. 

\bibitem[Risaliti et al.(2013)]{risaliti2013}
Risaliti, G., Harrison, F.~A., Madsen, K.~K., et al.\ 2013, \nat, 494, 449

\bibitem[Rogantini et al.(2022)]{rogantini2022}
Rogantini, D., Costantini, E., Gallo, L.~C., et al.\ 2022, \mnras, 516, 5171

\bibitem[{{Ross} \& {Fabian}(2005)}]{ross2005}
{Ross}, R.~R., \& {Fabian}, A.~C. 2005, \mnras, 358, 211

\bibitem[{{Sambruna} {et~al.}(2011){Sambruna}, {Tombesi}, {Reeves}, {Braito},
  {Ballo}, {Gliozzi}, \& {Reynolds}}]{sambruna2011}
{Sambruna}, R.~M., {Tombesi}, F., {Reeves}, J.~N., {et~al.} 2011, \apj, 734,
  105

\bibitem[Scheuer \& Feiler(1996)]{scheuer1996}
Scheuer, P.~A.~G. \& Feiler, R.\ 1996, \mnras, 282, 291

\bibitem[{{Shakura} \& {Sunyaev}(1973)}]{shakura1973}
{Shakura}, N.~I., \& {Sunyaev}, R.~A. 1973, \aap, 24, 337

\bibitem[{{Shannon}(1949)}]{shannon1949}
{Shannon}, C.~E. 1949, IEEE Proceedings, 37, 10

\bibitem[{{Sharma}(2017)}]{sharma2017}
{Sharma}, S. 2017, \araa, 55, 213

\bibitem[Silva et~al.(2018)]{silva2018}
Silva, C.~V., Costantini, E., Giustini, M., {et~al.} 2018, \mnras, 480, 2334

\bibitem[Sim et al.(2008)]{sim2008}
Sim, S.~A., Long, K.~S., Miller, L., et al.\ 2008, \mnras, 388, 611

\bibitem[{{Skilling}(2004)}]{skilling2004}
{Skilling}, J. 2004, in AIP Conf. Proc., Vol. 735, Bayesian Inference and Maximum Entropy Methods Science and Engineering: 24th International Workshop on Bayesian Inference and Maximum Entropy Methods in Science and Engineering, ed. R.~{Fischer}, R.~{Preuss}, \& U.~V. {Toussaint} (Melville, NY: AIP), 395

\bibitem[{{Sokal}(1996)}]{sokal1996}
{Sokal}, A.~D. 1996, Nuclear Physics B Proceedings Supplements, 47, 172

\bibitem[{{Storchi-Bergmann} {et~al.}(1995){Storchi-Bergmann}, {Eracleous},
  {Livio}, {Wilson}, {Filippenko}, \& {Halpern}}]{storchibergmann1995}
{Storchi-Bergmann}, T., {Eracleous}, M., {Livio}, M., {et~al.} 1995, \apj, 443, 617

\bibitem[{{Storchi-Bergmann} {et~al.}(1997){Storchi-Bergmann}, {Eracleous},
  {Teresa Ruiz}, {Livio}, {Wilson}, \& {Filippenko}}]{storchibergmann1997}
{Storchi-Bergmann}, T., {Eracleous}, M., {Teresa Ruiz}, M., {et~al.} 1997,
  \apj, 489, 87

\bibitem[{{Str{\"u}der} {et~al.}(2001){Str{\"u}der}, {Briel}, {Dennerl},
  {Hartmann}, {Kendziorra}, {Meidinger}, {Pfeffermann}, {Reppin}, {Aschenbach},
  {Bornemann}, {Br{\"a}uninger}, {Burkert}, {Elender}, {Freyberg}, {Haberl},
  {Hartner}, {Heuschmann}, {Hippmann}, {Kastelic}, {Kemmer}, {Kettenring},
  {Kink}, {Krause}, {M{\"u}ller}, {Oppitz}, {Pietsch}, {Popp}, {Predehl},
  {Read}, {Stephan}, {St{\"o}tter}, {Tr{\"u}mper}, {Holl}, {Kemmer}, {Soltau},
  {St{\"o}tter}, {Weber}, {Weichert}, {von Zanthier}, {Carathanassis}, {Lutz},
  {Richter}, {Solc}, {B{\"o}ttcher}, {Kuster}, {Staubert}, {Abbey}, {Holland},
  {Turner}, {Balasini}, {Bignami}, {La Palombara}, {Villa}, {Buttler},
  {Gianini}, {Lain{\'e}}, {Lumb}, \& {Dhez}}]{struder2001}
{Str{\"u}der}, L., {Briel}, U., {Dennerl}, K., {et~al.} 2001, \aap, 365, L18

\bibitem[Suebsuwong et al.(2006)]{suebsuwong2006}
Suebsuwong, T., Malzac, J., Jourdain, E., et al.\ 2006, \aap, 453, 773

\bibitem[Svoboda et al.(2012)]{svoboda2012}
Svoboda, J., Dov{\v{c}}iak, M., Goosmann, R.~W., et al.\ 2012, \aap, 545, A106

\bibitem[Tagliacozzo et al.(2023)]{tagliacozzo2023}
Tagliacozzo, D., Marinucci, A., Ursini, F., et al.\ 2023, \mnras, 525, 4735

\bibitem[{{Tanaka} {et~al.}(1995){Tanaka}, {Nandra}, {Fabian}, {Inoue},
  {Otani}, {Dotani}, {Hayashida}, {Iwasawa}, {Kii}, {Kunieda}, {Makino}, \&
  {Matsuoka}}]{tanaka1995}
{Tanaka}, Y., {Nandra}, K., {Fabian}, A.~C., {et~al.} 1995, \nat, 375, 659

\bibitem[Tarter et al.(1969)]{tarter1969}
Tarter, C.~B., Tucker, W.~H., \& Salpeter, E.~E.\ 1969, \apj, 156, 943

\bibitem[{{Tashiro} {et~al.}(2018)}]{tashiro2018}
{Tashiro}, M., {Maejima}, H., {Toda}, K., {et~al.} 2018, in \procspie, Vol. 10699, Space Telescopes and Instrumentation 2018: Ultraviolet to Gamma Ray, ed. J.-W.~A. {den Herder}, S.~{Nikzad}, \& K.~{Nakazawa}, 1069922

\bibitem[{{Thorne} \& {Price}(1975)}]{thorne1975}
{Thorne}, K.~S., \& {Price}, R.~H. 1975, \apjl, 195, L101

\bibitem[{{Torresi} {et~al.}(2010){Torresi}, {Grandi}, {Longinotti},
  {Guainazzi}, {Palumbo}, {Tombesi}, \& {Nucita}}]{torresi2010}
{Torresi}, E., {Grandi}, P., {Longinotti}, A.~L., {et~al.} 2010, \mnras, 401,
  L10

\bibitem[{{Tortosa} {et~al.}(2018){Tortosa}, {Bianchi}, {Marinucci}, {Matt}, \&
  {Petrucci}}]{tortosa2018}
{Tortosa}, A., {Bianchi}, S., {Marinucci}, A., {Matt}, G., \& {Petrucci}, P.~O. 2018, \aap, 614, A37

\bibitem[{{Turner} {et~al.}(2001){Turner}, {Abbey}, {Arnaud}, {Balasini},
  {Barbera}, {Belsole}, {Bennie}, {Bernard}, {Bignami}, {Boer}, {Briel},
  {Butler}, {Cara}, {Chabaud}, {Cole}, {Collura}, {Conte}, {Cros}, {Denby},
  {Dhez}, {Di Coco}, {Dowson}, {Ferrando}, {Ghizzardi}, {Gianotti}, {Goodall},
  {Gretton}, {Griffiths}, {Hainaut}, {Hochedez}, {Holland}, {Jourdain},
  {Kendziorra}, {Lagostina}, {Laine}, {La Palombara}, {Lortholary}, {Lumb},
  {Marty}, {Molendi}, {Pigot}, {Poindron}, {Pounds}, {Reeves}, {Reppin},
  {Rothenflug}, {Salvetat}, {Sauvageot}, {Schmitt}, {Sembay}, {Short},
  {Spragg}, {Stephen}, {Str{\"u}der}, {Tiengo}, {Trifoglio}, {Tr{\"u}mper},
  {Vercellone}, {Vigroux}, {Villa}, {Ward}, {Whitehead}, \&
  {Zonca}}]{turner2001}
{Turner}, M.~J.~L., {Abbey}, A., {Arnaud}, M., {et~al.} 2001, \aap, 365, L27

\bibitem[{{Ursini} {et~al.}(2018){Ursini}, {Petrucci}, {Matt}, {Bianchi},
  {Cappi}, {Dadina}, {Grandi}, {Torresi}, {Ballantyne}, {De Marco}, {De Rosa},
  {Giroletti}, {Malzac}, {Marinucci}, {Middei}, {Ponti}, \&
  {Tortosa}}]{ursini2018}
{Ursini}, F., {Petrucci}, P.~O., {Matt}, G., {et~al.} 2018, \mnras, 478, 2663

\bibitem[{{van den Bosch} {et~al.}(2015){van den Bosch}, {Gebhardt},
  {G{\"u}ltekin}, {Y{\i}ld{\i}r{\i}m}, \& {Walsh}}]{bosch2015}
{van den Bosch}, R. C.~E., {Gebhardt}, K., {G{\"u}ltekin}, K.,
  {Y{\i}ld{\i}r{\i}m}, A., \& {Walsh}, J.~L. 2015, \apjs, 218, 10

\bibitem[{{van der Walt} {et~al.}(2011){van der Walt}, {Colbert}, \&
  {Varoquaux}}]{walt2011}
{van der Walt}, S., {Colbert}, S.~C., \& {Varoquaux}, G. 2011, Computing in
  Science and Engineering, 13, 22


\bibitem[Walton et al.(2013)]{walton2013}
Walton, D.~J., Nardini, E., Fabian, A.~C., et al.\ 2013, \mnras, 428, 2901

\bibitem[{{Wilkins} \& {Fabian}(2011)}]{wilkins2011}
{Wilkins}, D.~R., \& {Fabian}, A.~C. 2011, \mnras, 414, 1269

\bibitem[{{Wilkins} \& {Fabian}(2012)}]{wilkins2012}
{Wilkins}, D.~R., \& {Fabian}, A.~C. 2012, \mnras, 424, 1284

\bibitem[Wilkins et~al.(2021)]{wilkins2021}
{Wilkins}, D.~R., {Gallo}, L.~C., {Costantini}, E., {Brandt}, W.~N., \&
  {Blandford}, R.~D. 2021, \nat, 595, 657

\bibitem[Wilkins et al.(2022)]{wilkins2022}
Wilkins, D.~R., Gallo, L.~C., Costantini, E., et al.\ 2022, \mnras, 512, 761

\bibitem[Wilks(1938)]{wilks1938}
Wilks, S.~S.\ 1938, Ann. Math. Stat., 9, 60

\bibitem[Wilks(1963)]{wilks1963}
Wilks, S.~S.\ 1963, Mathematical Statistics (Princeton: Princeton
University Press), Chapter 13

\bibitem[Williams et al.(2018)]{williams2018}
Williams, P.~R., Pancoast, A., Treu, T., et al.\ 2018, \apj, 866, 75

\bibitem[{{Wills} \& {Brotherton}(1995)}]{wills1995}
{Wills}, B.~J., \& {Brotherton}, M.~S. 1995, \apjl, 448, L81

\bibitem[{{Wilms} {et~al.}(2000){Wilms}, {Allen}, \& {McCray}}]{wilms2000}
{Wilms}, J., {Allen}, A., \& {McCray}, R. 2000, \apj, 542, 914

\bibitem[{{Wu} \& {Han}(2001)}]{wu2001}
{Wu}, X.-B., \& {Han}, J.~L. 2001, \apjl, 561, L59

\bibitem[Wu et al.(2022)]{wu2022}
Wu, Z., Ho, L.~C., \& Zhuang, M.-Y.\ 2022, \apj, 941, 95

\bibitem[{{Zdziarski} {et~al.}(1996){Zdziarski}, {Johnson}, \&
  {Magdziarz}}]{zdziarski1996}
{Zdziarski}, A.~A., {Johnson}, W.~N., \& {Magdziarz}, P. 1996, \mnras, 283, 193

\bibitem[{{Zhang} \& {Wu}(2002)}]{zhang2002}
{Zhang}, T.-Z., \& {Wu}, X.-B. 2002, \cjaa, 2, 487

\bibitem[{{Zhuang} {et~al.}(2018){Zhuang}, {Ho}, \& {Shangguan}}]{zhuang2018}
{Zhuang}, M.-Y., {Ho}, L.~C., \& {Shangguan}, J. 2018, \apj, 862, 118

\bibitem[{{{\.Z}ycki} {et~al.}(1999){{\.Z}ycki}, {Done}, \&
  {Smith}}]{zycki1999}
{{\.Z}ycki}, P.~T., {Done}, C., \& {Smith}, D.~A. 1999, \mnras, 309, 561

\end{thebibliography}

\end{document}